\documentclass[11pt, oneside]{article}   	
\usepackage{geometry}                		
\usepackage{graphicx}				
\usepackage{amssymb}
\usepackage{amsmath}
\usepackage{subfig}
\usepackage{caption}
\usepackage{comment}

\DeclareMathOperator{\arcsinh}{arcsinh}
\DeclareMathOperator{\arctanh}{arctanh}

\title{Dark Horse, Dark Matter: \\
Revisiting the $SO(16) \times SO(16)'$ Nonsupersymmetric Model in the LHC and Dark Energy Era}
\author{Michael McGuigan \\
Brookhaven National Laboratory}
\date{}							
\numberwithin{equation}{section}
\begin{document}
\maketitle

\begin{abstract}
We revisit the nonsupersymmetric $SO(16)\times SO(16)'$ model in light of LHC and Dark Energy data. Recently nonsupersymmetric models have become of great interest because the LHC has not found evidence of supersymmetry. In addition nonsupersymmetric models with a single Higgs-like field and small one loop  vacuum energy have been constructed. Also models of dark energy with a dilaton-radion  potential have also been recently examined in the light of dark energy data and the swampland conjecture. In this paper some of the features of the nonsupersymmetric $SO(16) \times SO(16)'$ model with regards to high energy physics and cosmology such as  dark energy, vacuum stablilization, dark matter candidates, dark matter portals, gauge-Higgs unification, and quantum cosmology are examined in the context of the  LHC and dark energy era.
\end{abstract}

\section{Introduction}

A dark horse is a little-known entity that emerges to prominence, especially in a competition of some sort that seems unlikely to succeed. The nonsupersymmetric $SO(16) \times SO(16)'$ model can be thought of as a dark horse, at least with respect to more popular $E_8 \times E_8'$, $Spin(32)/Z_2$ and Type II models.
The $SO(16)$ model was originally introduced as a model of flavor \cite{Wilczek:1981iz}\cite{Mohapatra:1979nn}
and was also studied in the Kaluza-Klein context as an example of a theory with gauge symmetry already in the higher dimensions and not coming from isometries in the compactified manifold \cite{Witten:1983ux}. Using the Atiya-Singer index theorem on this space one could then relate the number of families or chiral generations to one half the Euler characteristic of the manifold if one embeds the spin connection in the gauge group. 
The theory is part of a series of theories that can be considered with gauge group $SO(n)$ compactified on a space of dimension $n-10$ with fermions in the spinor representation $2^{\frac{n}{2} - 1}$ (See table [1]). For example $SO(12)$ in dimension six was considered in \cite{Wetterich:1985dn}\cite{Nomura:2008sx}\cite{Chiang:2011sj} and $SO(14)$ in eight dimensions was considered in  \cite{Jittoh:2009th}, $SO(10)$ in four dimensions is the usual GUT model \cite{Georgi:1979dq} and $SO(8)$ in two dimensions is realized in noncritical heterotic string theory \cite{McGuigan:1991qp}  .

\begin{table}[h]
\centering
\begin{tabular}{|l|l|l|l|}
\hline
Gauge Group        & Dimension  & Compact Space & Spinor Representation \\ \hline
$SO(18)$   &      12          & $S^8$   & 256    \\ \hline
$SO(16) $            & 10        & $S^6$           & 128                     \\ \hline
$SO(14)$          & 8                    & $S^4$  & 64                     \\ \hline
$SO(12)$            & 6                   & $S^2$   & 32                     \\ \hline
$SO(11)$ & 5                  & $S^1/Z_2$  & 32                      \\ \hline
$SO(10)$        & 4                    &  -  & 16                     \\ \hline
$SO(8)$        & 2                   &  -  & 8                     \\ \hline

\end{tabular}
\caption{\label{tab:table-name} Various $SO(n)$ models in dimension $(n-6)$ with fermions in the $2^{\frac{n}{2}-1}$ spinor representation.}
\end{table}

The UV complettion of the $SO(16) \times SO(16)'$ model is the unique nonsupersymmetric tachyon free string theory in ten dimensions. It was discovered by Dixon and Harvey \cite{Dixon:1986iz} and Alvarez-Gaume, Ginsparg, Moore and Vafa \cite{AlvarezGaume:1986jb} through studying a nonstandard projection of the string states in a way that was compatible with modular invariance and the absence of anomalies. The $SO(16)$ gauge group in ten dimensions was already seen as attractive in \cite{Green:1987mn} where dimensional reduction on a six dimensional manifold would lead to an $SO(10)$ Grand Unified group and a number of families of chiral fermions. Being nonsupersymmetric the theory has a non zero value for the cosmological constant in ten dimensions which is positive at the one loop level. This provides a way through compactification to reduce the four dimensional cosmological constant by balancing the large cosmological constant with the curvature of the extra dimensions. The nonzero comological constant also leads to dilaton tadpoles which can be consistently removed through the Fischler-Susskind mechanism. This has the effect of producing nonzero dilaton potentials and a dilaton mass as well as potentials and masses for other moduli fields associated with compactification. If the internal manifold is non simply connected with a discrete group that acts freely on the manifold one can further reduce the gauge group to products of Unitary groups similar to the standard model.

\begin{table}[h]
\centering
\begin{tabular}{|l|l|l|}
\hline
Internal manifold        & Euler characteristic & Number of chiral generations \\ \hline
$S^3 \times S^3$   &       0           & 0     \\ \hline
$S^5 \times S^1$   &       0           & 0     \\ \hline
$S^6 $            & 2                    & 1                     \\ \hline
$S^4\times S^2 $          & 4                    & 2                     \\ \hline
$CP^3$            & 4                    & 2                     \\ \hline
$Flag = SU(3)/U(1)^2$ & 6                    & 3                     \\ \hline
$(S^4 \times S^2) \# Flag \# (S^5 \times S^1)$ & 6                    & 3                     \\ \hline
$(S^2\times S^4) \#  (S^2\times S^4)$ & 6 & 3 \\\hline
$(S^2\times S^2 \times S^2) \#  (T^2\times S^2 \times S^2)$ & 6 & 3 \\\hline
$S^2\times S^2 \times S^2$        & 8                    & 4                     \\ \hline
\end{tabular}
        \caption{\label{tab:table-name}Simple compactified spaces of the $SO(16) \times SO(16)'$ nonsupersymmetric model as well as the Euler characteristic and four dimensional number of chiral generations or families. The notation $\#$ denotes topological connected sum.}
\end{table}

The internal manifolds considered for the compactification of the $SO(16) \times SO(16)'$ nonsupersymmetric string can be much simpler than those used for the supersymmetric strings see table [2]. This is because the internal manifold does not require Ricci flatness when the four dimensional effective field theory is nonsupersymmetric. The recent data form the LHC also supports the idea that nature may be nonsupersymmetric as all experiments to date are consistent with the nonsupersymmetric standard model. In addition  decays which proceed through loop effects are sensitive to supersymmetry and these also are consistent with the nonsupersymmetric standard model. While it is too early to rule out low energy supersymmetry the current data from LHC experiment is setting constraints on the parameter space associated with these theories. Thus from the point of view of simplicity of the model's internal manifolds as well as the absence of supersymmetry in early experimental results one can revisit the $SO(16)\times SO(16)'$ nonsupersymmetric string as a alternative starting point for an underlying fundamental theory of gravity and matter. Indeed investigations exist using nonsupersymmetric compactifications on Coset manifolds \cite{Lust:1986kj}\cite{Chapline:1983zh}\cite{Hanlon:1992mn}\cite{Kapetanakis:1990xx}, tori \cite{Ginsparg:1986wr}\cite{Nair:1986zn}, orbifolds\cite{Font:2002pq}\cite{Abel:2015oxa} and Calabi-Yau manifolds \cite{Blaszczyk:2014qoa} \cite{Blaszczyk:2015zta} have been developed and are being intensively pursued. Nonperturbative approaches of the $SO(16)\times SO(16)'$ model have been considered using a Melvin background in M-theory \cite{Suyama:2001ne}\cite{Motl:2001dj}, additional projections of states in Horava-Witten theory \cite{Faraggi:2007tj}, Type I interpolating models \cite{Blum:1997cs}\cite{Blum:1997gw} and realization of twisted states from orbifolds  in M-theory\cite{Kaplunovsky:1999ia}\cite{Gorbatov:2001pw}\cite{Claussen:2016ucd}. Without supersymmetric nonrenormalization theorems it is difficult to make precise statements about nonsupersymmetric dual theories at  strong coupling however \cite{Banks:1999tr}.

Criticism of string theory in general includes failure of convergence of the perturbative series \cite{Gross:1988ib}, nonlocality and instability from higher derivative interactions \cite{Eliezer:1989cr}, lack of predictions, overly complicated models and lack of a compelling underlying theory\cite{Davies:1988dm}\cite{Ginsparg:1986an}\cite{Woit:2006js}\cite{Hossenfelder:2018}. For nonsupersymmetric string theory one also has instability with respect to transition back to a supersymmetric theory and presence of dilaton tadpoles\cite{Zichichi:1990ui}. For most of this paper however except for discussion in section 2  we will use low energy effective field theories instead of the UV complete string theory.

This paper is organized as follows. In section 2 we recall some aspects of  the $SO(16) \times SO(16)'$ string, it's construction, spectrum including a bi-fundamental fermion, and 1-loop cosmological constant. In section 3 we discuss compactifications of the model and how the large ten dimensional 1-loop cosmological constant can be reduced through compactification. We derive effective potentials for the radion as a function of internal flux and also discuss the stabilization of the dilaton which has a runaway potential in ten dimensions. In section 4 we discuss dark matter candidates for the $SO(16) \times SO(16)'$ nonsupersymmetric model and the effect of the model's bifundamental fermion as a portal, mediator or connector field into the dark sector and on dark matter production in the form of dark glueballs. In section 5 we discuss how the Higgs field is realized in the model and the implications for Higgs physics including Higgs decays and the effects on the Higgs potential from the bifundamental fermion. In section 6 we discuss cosmology in the $SO(16)\times SO(16)'$ nonsupersymmetric theory including cosmologies with the dark gauge field. In section 7 we discuss the Fischler-Susskind mechanism and the introduction of nonperturbative dilaton potentials into the model. In section 8 we discuss the quantum cosmology of a dimensionally reduced version of the theory. Finally in section 9 we discuss the main conclusions of the paper.

\section{Review of $SO(16)\times SO(16)^\prime$ nonsupersymmetric string}

The construction of the $SO(16)\times SO(16)^\prime$ nonsupersymmetric string is described in detail in \cite{Green:1987mn}. There are two equivalent formulations of the theory as a twisted version of the heterotic string.
In the bosonic formulation of the $SO(16) \times SO(16)'$ model one considers twists of the form:
\begin{equation}R={\left. {{e^{2\pi i{J_{12}}}}} \right|_{spacetime}} \cdot {\gamma _\delta }\end{equation}
Here the first factor is  an operator which generates a $2 \pi$ rotation and $\gamma _\delta $ is a translation in the root lattice of $E_8 \times E_8$ by $\pi \delta$. For $\delta^2 = 2$ there are no tachyons in the theory. The choice for $\delta$ is given by:
\begin{equation}\delta  = (1,0,0,0,0,0,0,0,1,0,0,0,0,0,0,0)\end{equation}
which reduces the root lattice to $D_8 \times D_8$, the root lattice of $SO(16)\times SO(16)'$
In the fermionic formulation of the $SO(16)\times SO(16)'$ theory one represents the internal coordinates by fermions and the the $\gamma _ \delta$ operator  by the product of two rotations represented by these fermions. The $\gamma_\delta$ operator is written as:
\begin{equation}R = {\left. {{e^{2\pi i{J_{12}}}}} \right|_{spacetime}} \cdot {e^{2\pi i{j_{12}}}} \cdot {e^{2\pi ij{'_{12}}}}\end{equation}
Here $j_{12}$ and $j'_{12}$ generate rotations in $SO(16)$ and $SO(16)'$.
In either formulation one obtains physical states which survive these projections and massless states given in table [3] including twisted states containing a massless right handed fermion that are necessary to realize modular invariance in the theory.

\begin{table}[h]
\centering
\begin{tabular}{|l|l|l|}
\hline
Fermions     & Description & $4096$ degrees of freedom  \\ \hline
$\psi$   &     visible fermion & $(128)8 = 1024 $          \\ \hline
$\psi'$         & hidden fermion  & $(128)8=1024$ \\ \hline
$\chi$        & bi-fundamental portal fermion  & $(256)8 = 2048$                        \\ \hline
    Bosons      & Description  &  $1984$   degrees of freedom \\ \hline
$A_{\mu}$ & visible gauge boson - visible gluon& $120(8)= 960 $                  \\ \hline
$A_{\mu}'$ & hidden gauge boson - dark gluon & $120(8)= 960 $                  \\ \hline
$g_{\mu\nu}$ & metric - graviton & $35 $                  \\ \hline
$B_{\mu\nu}$  & antisymmetric tensor field - axion      & $28$                    \\ \hline
$\phi$      & dilaton  - potential inflaton & $1$                  \\ \hline
\end{tabular}
\caption{\label{tab:table-name}Massless field content of the $SO(16) \times SO(16)'$ nonsupersymmetric model. Note the model has no gauged scalars in ten dimensions, a hidden $SO(16)'$ sector and a bi-fundamental fermion field $\chi$ connecting the two gauged sectors. Also note that the theory has an excess of fermions over bosons that leads to a positive one-loop ten dimensional cosmological constant.}
\end{table}

The massless sector of the $SO(16)\times SO(16)'$ nonsupersymmetric string contains the usual metric, dilaton and antisymmetric tensor fields as well gauge bosons in the adjoint representation $(120,1)$ and $(1,120')$ and chiral ferimons in the spinor $(128,1)$ and $(1,128')$ representation as well as bi-fundamental fermions of the opposite chirality in the co-spinor $(16,16')$ representation of $SO(16)\times SO(16)'$. Notably one has an absence of scalar transforming under a representation of the gauge group.

The four point tree level amplitude for the $O(16)\times O(16)$ nonsupersymmetric String was constructed in \cite{Kostelecky:1986ub}. Although branch cuts are seen in operator product expansions the full amplitude has the right physical behavior for a four fermion scattering amplitude with no tachyon pole.

\subsection{Ten dimensional effective action}

The Chapline-Manton type action for the ten dimensional massless fields takes the form $S_{10}=S_{dil-grav} + S_{antisym} + S_{gauge} + S_{fermion}$
where
$$
{S_{dil-grav}} =   \frac{1}{{2{\kappa ^2}}}\int {{d^{10}}x} \sqrt { - g} {e^{ - 2\phi }}(R + 4{(\nabla \phi )^2}  - 2{e^{2\phi }}{\lambda _{10}})\\
$$
$$
{S_{antisym}} =   \frac{1}{{2{\kappa ^2}}}\int {{d^{10}}x} \sqrt { - g} {e^{ - 2\phi }}(-\frac{1}{2} H^2 ))\\
$$
\begin{equation}
{S_{gauge}} =   \frac{1}{{2{\kappa ^2}}}\int {{d^{10}}x} \sqrt { - g} {e^{ - 2\phi }}(-\frac{1}{2}(F^2+F'^2))\\
\end{equation}
Finally adding the fermions to the effective action adds the $S_{fermion}$ to the action where:
\begin{equation}
{S_{fermion}} = \frac{1}{{2{\kappa ^2}}}\int {{d^{10}}z{e^{-2 \phi}}\det (e)({e^{ - 1}})_A^M(\bar \psi } {\gamma ^A}{D_M}\psi  + \bar \psi '{\gamma ^A}{D_M}\psi ' + \bar \chi {\gamma ^A}{D_M}\chi )\\
\end{equation}
and covariant derivatives are defined by:
$$
{D_M}\psi  = {\partial _M}\psi  + i{A_M}\psi 
$$
$$
{D_M}\psi ' = {\partial _M}\psi ' + iA{'_M}\psi '
$$
\begin{equation}
{D_M}\chi  = {\partial _M}\chi  + i{A_M}\chi  - i\chi A{'_M}
\end{equation}
The portion of the effective action $S_{gauge} + S_{fermion}$ will be important when we discuss the implications of the model for dark matter as it will contain dark matter candidates as well as portal or connector fields that connect the the visible sector to the dark hidden sector.

\subsection{ Calculation of $\lambda_{10}$}

In this subsection we recall the calculation of the one-loop cosmological constant for the $SO(16) \times SO(16)'$ nonsupersymmetric model first done in \cite{AlvarezGaume:1986jb}. The calculation of $\lambda_{10}$ involves expanding out the integrand of bosonic and fermionic partition functions $P_B$ and $P_F$ of the model in a power series in ${q^m}{{\bar q}^n}$ and integrating term by term. 

For the $SO(16) \times SO(16)'$ nonsupersymmetric model we have:
$${P_B} = {2^9}\frac{1}{{\theta _1^{'4}\bar \theta _1^{'8}}}\{ \theta _2^4({(\bar \theta _3^8 + \bar \theta _4^8)^2} + \bar \theta _2^{16}) + (\theta _3^4 + \theta _4^4){(\bar \theta _3^8 - \bar \theta _4^8)}2\bar \theta _2^8\} $$
\begin{equation}{P_F} = {2^9}\frac{1}{{\theta _1^{'4}\bar \theta _1^{'8}}}\{ \theta _2^4({(\bar \theta _3^8 - \bar \theta _4^8)^2} + \bar \theta _2^{16}) + (\theta _3^4 - \theta _4^4){(\bar \theta _3^8 + \bar \theta _4^8)}2\bar \theta _2^8\} \end{equation}
In the notation of \cite{Abel:2015oxa} the one loop cosmological constant is:
\begin{equation}{\lambda _{10}} =\frac{1}{2}\frac{1}{{\alpha {'^5}}}\frac{1}{{{{(2\pi )}^{10}}}}\int\limits_F {\frac{{{d^2}\tau }}{{\tau _2^6}}} ({P_F} - {P_B})\end{equation}
Expanding out the integrand as a Laurent series in $q$ and $\bar q$ we define the coefficients $f_{m,n}$ as:
\begin{equation}({P_F} - {P_B}) = 2^7\sum\limits_{m,n} {{f_{m,n}}{q^m}{{\bar q}^n}} \end{equation}
The difference in fermion and boson partition function was computed in \cite{Dixon:1986iz} \cite{AlvarezGaume:1986jb} and  is written in terms of Jacobi theta functions as:
\begin{equation}({P_F} - {P_B}) = -{2^7}\frac{1}{{\theta _1^{'4}}}\left( {\frac{{\theta _2^4}}{{\bar \theta _2^8}} + \frac{{\theta _4^4}}{{\bar \theta _4^8}} - \frac{{\theta _3^4}}{{\bar \theta _3^8}}} \right)\end{equation}
Expanding this function out we have:
\begin{equation}
(P_F-P_B)(q,\bar{q}) = 2^74 \left(9207 q^2 \bar{q}^2-\frac{q^2}{8 \bar{q}^2}-\frac{480 \bar{q}^3}{q}-288 q
   \bar{q}-\frac{8 \bar{q}}{q}-\frac{1}{64 \bar{q}^2}+\frac{10263
   q^2}{16}+\frac{33}{8}+ \ldots \right)
   \end{equation}
As in \cite{Abel:2015oxa} we can define integrals by:
\begin{equation}{I_{m,n}} = \int\limits_F {{d^2}\tau \frac{1}{{\tau _2^6}}} {q^m}{{\bar q}^n}\end{equation}
These integrals are given numerically by:
\begin{equation}\begin{array}{l}
{I_{0, - 2}} =  - 14.258\\
{I_{ - 1,1}} =  - .038\\
{I_{0,0}} = .257\\
{I_{2, - 2}} = .014\\
{I_{1,1}} = 3.021 \times {10^{ - 4}}\\
{I_{ - 1,3}} = 4.682 \times {10^{ - 5}}\\
{I_{2,2}} = 5.7591 \times {10^{ - 7}}\\
{I_{2,0}} =  - 1.029 \times {10^{ - 4}}
\end{array}\end{equation}
Now from (2.9) as:
\begin{equation}\begin{array}{l}
{f_{0, - 2}} =  - 1/16\\
{f_{ - 1,1}} =  - 32\\
{f_{0,0}} = 33/2\\
{f_{2, - 2}} = -1/2\\
{f_{1,1}} = -4(288) \\
{f_{ - 1,3}} = -4(480) \\
{f_{2,2}} = 4(9207) \\
{f_{2,0}} =  10263/4
\end{array}\end{equation}
Now defining $\lambda_{m,n} = f_{m,n}I_{m,n}$ the contribution from each term is:
\begin{equation}\begin{array}{l}
{\lambda _{0, - 2}} = .891125\\
{\lambda _{ - 1,1}} = 1.216\\
\lambda {}_{0,0} = 4.2405\\
{\lambda _{2, - 2}} =  - .007\\
{\lambda _{1,1}} =  - .348019\\
{\lambda _{ - 1,3}} = .0898944\\
{\lambda _{2,2}} = .0212096\\
{\lambda _{2,0}} =  - .264016
\end{array}\end{equation}
So that the total expression for the one loop cosmological constant is given by:
\begin{equation}{\lambda _{10}} = {\alpha'}^{-5}{2^6}{(2\pi )^{ - 10}}(\sum\limits_{m,n} {{\lambda _{m,n}})}  = {{\alpha '}^{ - 5}}{2^6}{(2\pi )^{ - 10}}(5.65991)\end{equation}
so that 
\begin{equation}\lambda_{10} = 3.77738 \times 10^{-6}{{\alpha '}^{ - 5}}\end{equation}
in agreement with \cite{AlvarezGaume:1986jb}\cite{Hamada:2015ria}. In the following we will chose units so that $\alpha' =1$. The effective four dimensional Newtons constant can be determined by $\frac{1}{{16\pi G}} = \frac{1}{2}\frac{1}{{\alpha {'^4}}}{V_6}{e^{ - 2{\phi _0}}}$ where $V_6$ is a compactified volume and $\phi_0$ the zero mode of the dilaton.
Note that if we pull out the contribution from the massless modes this can be written using 
$
I_{0,0} = \frac{4}{9 \sqrt{3}}
$
and 
$
(n_F^0 - n_B^0)=2^7f_{0,0}    
$
as:
\begin{equation} {\lambda _{10}} = \frac{1}{2}\frac{1}{{\alpha {'^5}}}\frac{1}{{{{(2\pi )}^{10}}}}((n_F^0 - n_B^0)\frac{4}{{9\sqrt 3 }} + {c_{r}}) \end{equation}
with $(n_F^0 - n_B^0)= 2112$ for the massless modes of the  $SO(16)\times SO(16)^\prime$ nonsupersymmetric string and ${c_{r}}= 182.529$. The massless mode contribution to $\lambda_{10}$ is about three times the contribution from $c_r$. 
One can follow a similar procedure to calculate one loop cosmological constants for nonsupersymmetric orbifold compactifications to four dimensions. In some cases when $ n_{F}^0-n_{B}^0=0 $  one can can obtain an exponentially suppressed four dimensional cosmological constant \cite{Itoyama:1987rc}\cite{Itoyama:1986ei}\cite{Abel:2015oxa}\cite{Itoyama:2019yst}. Under additional conditions one can also obtain a suppressed one loop correction to  scalar masses for these theories as well\cite{Abel:2015oxa}.

\section{ Dark energy in the Compactified $SO(16) \times SO(16)'$ nonsupersymmetric model }

Although the $SO(16) \times SO(16)'$ nonsupersymmetric model  is of some interest as it is an example a string model with positive cosmological constant the one loop generated dilaton potential is a runaway exponential or Liouville-type potential. These type of potentials are in tension with astrophysical data \cite{Akrami:2018ylq}. Also the value of the one-loop cosmological constant is far too large to constitute the dark energy. Nevertheless compactifications of theories with positive cosmological constant are of interest especially for flux compactifications \cite{Kolb:1986nj}\cite{Carroll:2009dn}\cite{Maloney:2002rr}\cite{Das:2019vnx} which can effectively reduce the value of the ten dimensional cosmological constant. To analyze these flux compactifications We will proceed in two stages. First we will look at flux compactifications with a frozen dilaton just considering the radion and flux parameters. In the second stage we will discuss what happens when one turns on the dilaton field in the presence of modified dilaton potentials.

\subsection{Einstein-Maxwell flux compactifications of the  $SO(16) \times SO(16)'$ nonsupersymmetric model }

There are two types of flux compactifications we can consider for the  $SO(16) \times SO(16)'$ nonsupersymmetric model. One can consider two form fluxes from Abelian components of the $F$ field or three form fluxes associated with the antisymmetric $H$ field. In this paper we consider the former with Abelian components arising through the embedding $SO(16)  \supset SU(8) \times U(1)$ with the Abelian two form flux associated with the $U(1)$ factor. We consider the compact manifold $S^2 \times S^2 \times S^2$ which is of the form of the Einstein-Maxwell landscape considered in \cite{Brown:2014sba}\cite{Brown:2013mwa}\cite{Brown:2013fba}\cite{Kan:2015cia}\cite{Hertzberg:2015bta}\cite{Ramadhan:2015ona}\cite{Asensio:2012wt}\cite{Asensio:2012pg}\cite{Dienes:2004pi}.

\subsection{Compactification of the non-supersymmetric string and effective potential}

One can consider compactifications of the $SO(16)\times SO(16)^\prime$ nonsupersymmetric string to four dimensions. Compactifcations on Tori, orbifolds and Calabi-Yau manifolds are can be considered. Unlike the supersymmetric case the internal manifolds need not be Ricci flat.  As in the supersymmetric case the number of fermion generations is given by half the Euler characteristic if one embeds the spin connection in the gauge group. For the internal manifold $S^2 \times S^2 \times S^2$  which will allow us to apply recent results from studies of the Einstein-Maxwell landscape. We will denote the flux squared by $f^2 = F_1^2 + F_2^2 + F_3^2 + {F'}_1^2 + {F'}_2^2 + {F'}_3^2$ where $F_i$ and $F'_i$ are the flux through the $i$th two sphere in $S^2 \times S^2 \times S^2$.

Assuming the radius of each $S^2$ is given by $b$ the total volume of the internal six dimension space is $(4 \pi )^3 b^6$ and the Ricci scalar curvature of the six dimensional internal space is $3 (2 b^{-2})$.
The effective four dimensional Lagrangian is then:
 $$ \frac{1}{2} (4 \pi )^3\int {{d^4}x} \sqrt { - g} {e^{ - 2\phi }}{b^6}(R + 4{(\nabla \phi )^2} - 12\frac{{\square b}}{b} -30\frac{{{{(\nabla b)}^2}}}{{{b^2}}} + 6\frac{1}{{{b^2}}}  $$
  \begin{equation}  - 2 {\lambda _{10}}{e^{2\phi }}  -  \frac{{f^2}}{{{b^{4}}}}) \end{equation}
To remove the second derivative one can integrate the third term by parts to obtain:
$$\frac{1}{2} (4 \pi )^3\int {{d^4}x} \sqrt { - g} {e^{ - 2\phi }}{b^6}(R + 4{(\nabla \phi )^2} + 30\frac{{{{(\nabla b)}^2}}}{{{b^2}}} - 24\nabla \phi  \cdot \frac{{\nabla b}}{b} + 6 \frac{1}{{{b^2}}}$$
\begin{equation} -2{\lambda _{10}}{e^{2\phi }}  -  \frac{{f^2}}{{{b^{4}}}})\end{equation}
To transform to the Einstein frame we use the Weyl transformation ${\Omega ^2}{g_{\mu \nu }}$ such that ${e^{ - 2\phi }}{b^6}{\Omega ^{2}} = 1$ and $\Omega  = {e^{\phi }}{b^{ - 3}}$. The Weyl transformed Lagrangian is then written as:
$$\frac{1}{2} (4 \pi )^3\int {{d^4}x} \sqrt { - g} (R - 2{(\nabla \phi )^2} + 12\nabla \phi  \cdot \frac{{\nabla b}}{b} - 24\frac{{{{(\nabla b)}^2}}}{{{b^2}}}$$
\begin{equation} + {e^{2\phi }}{b^{ - 6}}(6\frac{1}{{{b^2}}} - 2 {\lambda _{10}}{e^{2\phi }}  -  \frac{{f^2}}{{{b^{4}}}}) )\end{equation}
We can then write the compactified action in the form:
\begin{equation}\frac{1}{2} (4 \pi )^3\int {{d^4}x} \sqrt { - g} (R - 2{(\nabla \phi )^2} + 12\nabla \phi  \cdot \frac{{\nabla b}}{b} - 24\frac{{{{(\nabla b)}^2}}}{{{b^2}}} - 2 V_{eff}(b,\phi))\end{equation}
where:
\begin{equation}V_{eff}= (4 \pi)^3 {e^{2\phi }}{b^{ - 6}}(-3\frac{1}{{{b^2}}} + {\lambda _{10}}{e^{2\phi }}  + \frac{1}{2}\frac{{f^2}}{{{b^{4}}}}) \end{equation}
This form will be useful when we study the the relation of dark energy to the non-supersymmetric string.

\subsection{Effective potential with curvature, radion and flux}

The effective potential for the radion after Weyl rescaling for the radion to go to the Einstein frame is then given by:
\begin{equation}V_{eff}(b) = {(4\pi )^3}{b^{ - 6}}( - 3{b^{ - 2}} + {\lambda _{10}} + \frac{1}{2}{f^2}{b^{ - 4}})\end{equation}
Here we simplified the potential by assuming all the radii and flux of the each of the $S^2$ are equal with the dilaton frozen. 
The first term in the potential is related to the curvature of the compactified space, the $\lambda_{10}$ term is from the one loop ten dimensional cosmological constant and the final term is the flux contribution.  The condition for a local minimum are
\begin{equation}\frac{{d{V_{eff}}({b_m})}}{{db}} = 0\end{equation}
with the value of the effective cosmological constant given by the value of the effective potential at the local minimum given by:
\begin{equation}\lambda_4 = V_{eff}({b_m})\end{equation}
We plot the potential for $f=1092$ in figure 1. The potential has a similar shape to those found in other studies of compactified theories with a higher dimensional cosmological constant
\cite{Kolb:1986nj}
  \cite{Carroll:2009dn}
\cite{Maloney:2002rr}
\cite{Das:2019vnx}.
The potential has a local minimum at at $b_m = 630.93$ and a reduced effective vacuum energy $\lambda_{4} = 1.16061 \times {10^{ - 22}}$ so DeSitter space with a reduced cosmological constant in four dimensions.
Raising $f$ above $1127$ and the local minima becomes a saddle point figure 2 ,while lowering $f$ one can obtain a negative value for $\lambda_{4}$ or Anti-DeSitter space.  The solution for the flux parameter $f_0$ and internal radius $b_0$ corresponding to four dimesnional Minkowski space are given by:
$${b_0} = \sqrt {\frac{3}{2}} \frac{1}{{\sqrt {{\lambda _{10}}} }} = 630.15921$$
\begin{equation}{f_0} = \sqrt 3 {b_0} = \frac{3}{{\sqrt 2 }}\frac{1}{{\sqrt {{\lambda _{10}}} }} = 1091.46778\end{equation}
The relations are similar to other studies of compactified extra dimensions with a higher dimensional cosmological constant and gauge fields
\cite{RandjbarDaemi:1982hi}
\cite{Kubyshin:1993yc}
\cite{Surridge:1987ws}.

\begin{figure}
  \includegraphics[width =  \linewidth]{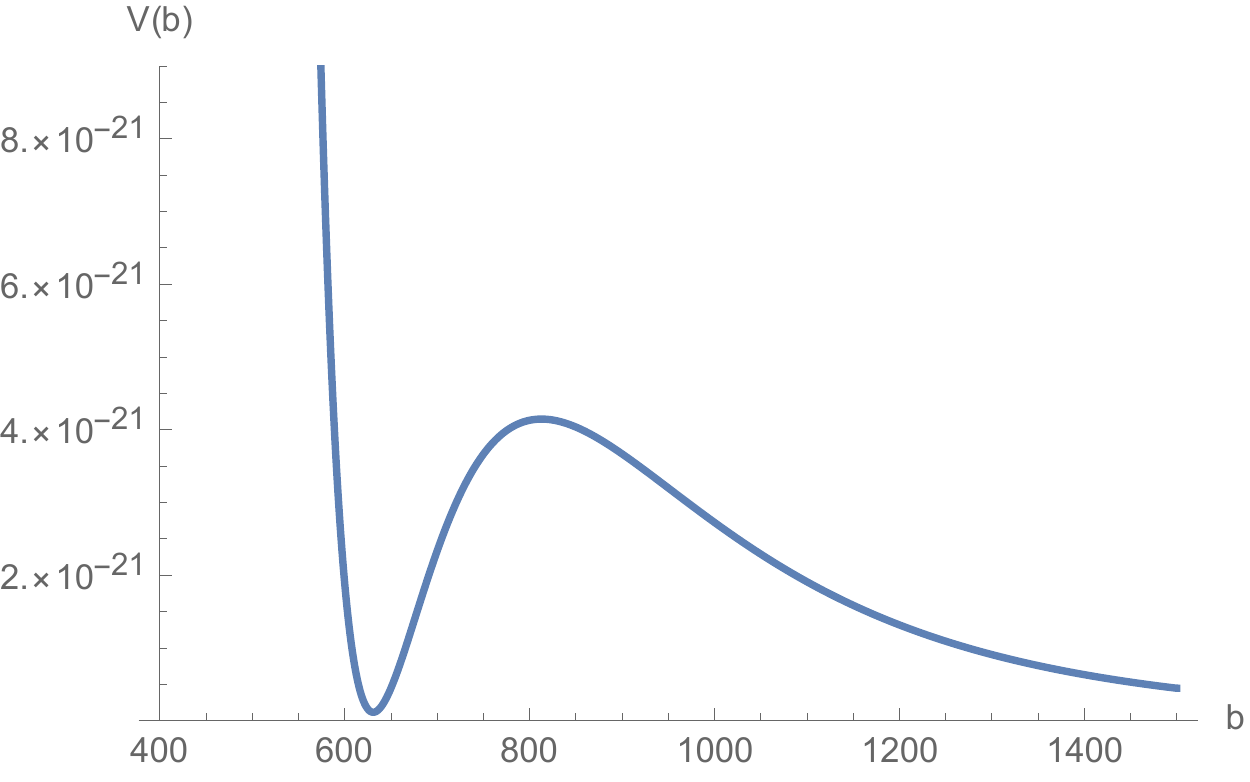}
  \caption{Radion potential for flux $f=1092$. The potential admits a local minimum at $b_0 = 630.93$ and effective positive four dimensional cosmological constant $\lambda_{eff} = 1.16061 \times {10^{ - 22}}$.}
  \label{fig:Radion Potential}
\end{figure}

\begin{figure}%
\centering
\subfloat[a]{%
\label{fig:first}%
\includegraphics[height=1.5in]{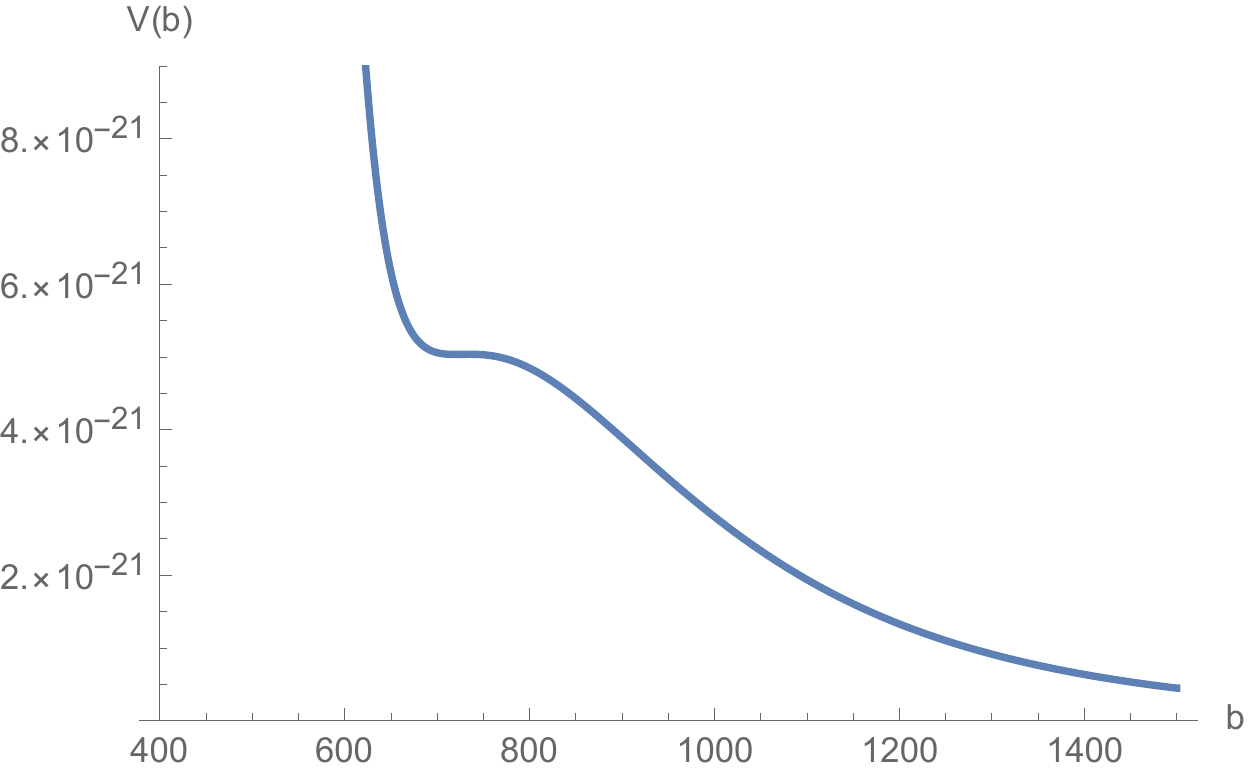}}%
\qquad
\subfloat[b]{%
\label{fig:second}%
\includegraphics[height=1.5in]{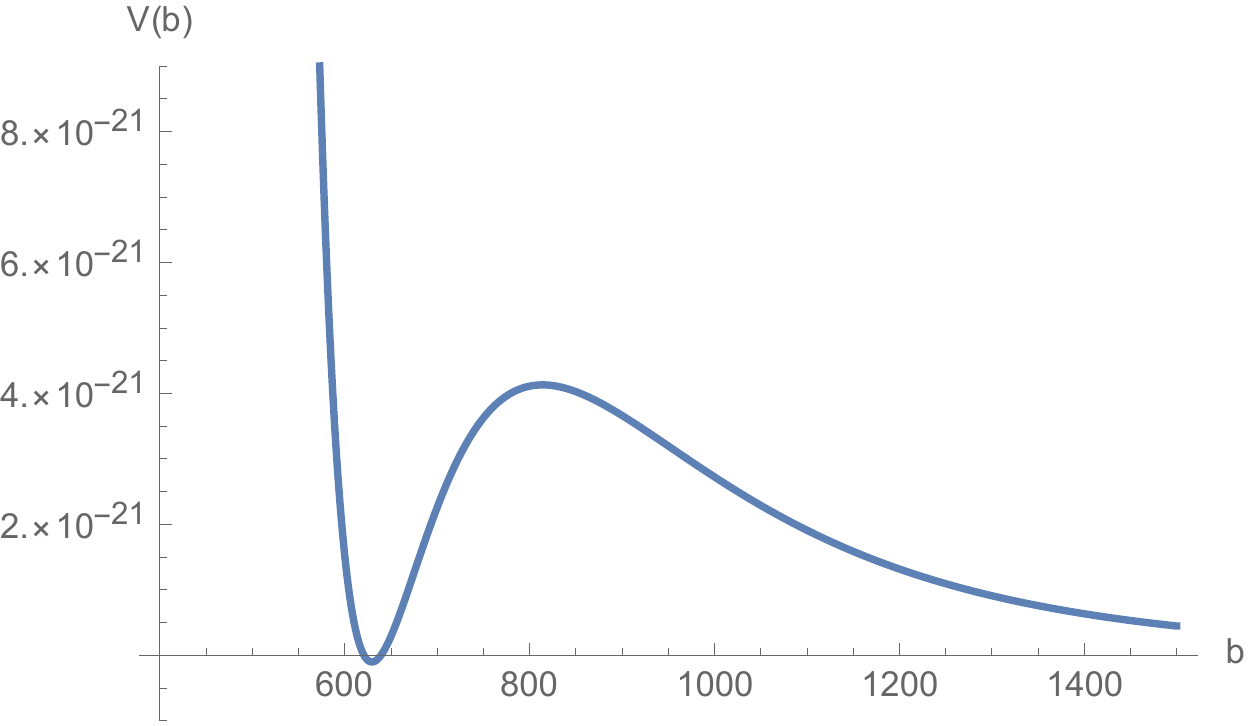}}%
\caption{[a] Radion potential for flux $f=1127$. The potential admits a saddle point  with a hilltop type potential. [b] Radion potential for flux $f=1091$. The potential admits an effective negative four dimensional cosmological constant.}
\end{figure}

Thus with the dilaton frozen it is possible to obtain a positive four dimensional cosmological constant in the compactified $SO(16)\times SO(16)'$ nonsupersymmetric model. Note however this is a local minimum and represents a meta-stable vacuum. Eventually the Universe will tunnel to large values of $b$ and decompactify. This is counterintuitive as one is used to thinking that higher dimensions manifest themselves only at early times and short distances while in this case if one waits long enough the theory decompactifies to it's ten dimensional origins. The eventual decay of the metastable state is an example of spontaneous decompactification \cite{Giddings:2004vr}. The model can undergo spontaneous compactification to four dimensions and the then spontaneous decompactification back to ten dimensions. This is consistent with what we see if the tunneling probability for decompactification is low enough. Finally decompactification could also be relevant for the very early Universe where all in the dimensions start out curled up and then some of them decompactify \cite{Brandenberger:1988aj}. For example one can consider initial states with the dimensions curled up in a group manifold like $U(3)$ or $SU(2)^3$ and decompactify to $SU(3)/U(1)^2 \times T^3$ or $S^2 \times S^2 \times S^2 \times T^3$ respectively with the $T^3$ radius taken to be large.

\subsection{Effective potential with curvature, dilaton, radion and flux}

To avoid the dilaton runaway potential one can consider adding an additional dilaton stabilization potential $V_{np}(\phi)$. Then the dilaton-radion potential after Weyl rescaling to the Einstein frame takes the form:
$$V_{eff}(b,\phi ) = {(4\pi )^3}{b^{ - 6}}{e^{2\phi }}( - 3{b^{ - 2}} + {\lambda _{10}}{e^{2\phi }} + \frac{1}{2}{f^2}{b^{ - 4}}) + {V_{np}}(\phi )$$
The potential for zero stabilization potential is shown in figure 3. The potential has a local minimum at negative effective four dimensionaal cosmological constant which is consistent with the swampland conjecture. The conditions for a local minimum are:

\begin{equation}\left| {\nabla {V_{eff}}({b_m},{\phi _m}} \right| = 0\end{equation}

Then the effective four dimensional cosmological constant will be the value at the local minimum given by:
\begin{equation}\lambda_4 = V_{eff}({b_m},{\phi _m})\end{equation}

Introducing a nonzero dilaton stabilization potential can allow for the possibility of a positive effective four dimensional cosmological constant. The dilaton stabilization potential could have several origins but is most likely of non-perturbative origin. In Horava-Witten theory the dilaton is interpreted as the radius of a eleventh dimension and the stabilization potential could be generated as a Casimir potential with respect to extra compact dimension as a has been done of the Heterotic Horava-Witten theory in \cite{Moss:2011pi}\cite{Fabinger:2000jd}\cite{Sakamura:2010ju}
\cite{Teo:2009bv}\cite{Ahmed:2009ty}\cite{Buchbinder:2003pi}\cite{Garriga:2000jb}. Another approach to obtain a dilaton stabilization potential is through fermion condensates in the hidden sector in a similar manner to the gluino condensates in the supersymmetric Heterotic string. Some approaches to  condensates in the $SO(16)$ model were considered in \cite{Hanlon:1993aq}. One can also add a stabilization potential explicitly as was done in by Horowitz and Horne and Gregory and Harvey \cite{Horne:1992bi}\cite{Gregory:1992kr}. This has the effect of avoiding the Brans-Dicke type gravity model associated with the massless dilaton by giving the dilaton a mass. This is consistent with precision gravity tests and as there is no local gauge principle associated with the dilaton (unlike the graviton and photon) and there is no inconsistency in giving the dilaton mass. Horowitz and Horne considered two types of stabilization potentials one proportional to ${(\sinh (2\phi ))^2}$ and another proportional to ${\phi ^2}$. 

In figure 4 we show the dilaton-radion potential associated with ${(\sinh (2\phi ))^2}$  and the dilaton-radion potential associated with the ${\phi ^2}$ stabilization potential. In both cases it was possible to find a local minimum with positive effective four dimensional cosmological constant. This is because the dilaton stabilization potential has the effect of confining the dilaton value to a narrow range, effectively freezing out the dilaton field and reducing the analysis to the flux potential without dilaton described above. This is still consistent with the swampland conjecture though as the dilaton stabilization potential was put in by hand rather than being generated by string theory. One can pursue the Horava-Witten theory Casimir approach or the nonperturbative fermion condensate approach to dilaton stabilization to investigate  a self-contained mechanism to obtain positive effective four dimensional cosmological constant or DeSitter space in the  $SO(16) \times SO(16)'$ nonsupersymmetric model. 

By allowing for different values for the fluxes and internal radii of the $S^2$'s one can obtain an effectively fractional $f_0$ and reduce to effective four dimensional cosmological constant further in a manner similar to the mechanism of Bousso-and Polchinski with four form fluxes. For example for $f_0=1091.46777$ one can obtain effective four dimensional constants as small as $2 \times 10^{-38}$. Using numerical algorithms discussed by Bao, Bousso, Jordan and Lackey\cite{Bao:2017thx} one can even further reduce the effective four dimensional cosmological constant value above zero.

\begin{figure}
  \includegraphics[width =  \linewidth]{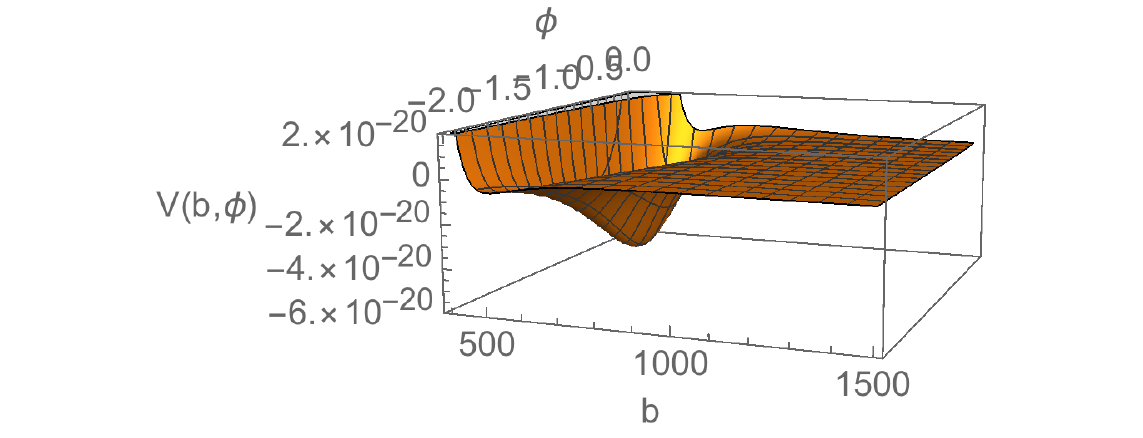}
  \caption{Dilaton-Radion potential for flux $f=1092$ and zero dilaton stabilization potential. The potential admits a local minimum with negative  effective four dimensional cosmological constant which is consistent with the swampland conjecture.}
  \label{fig:Dilaton-Radion Potential with zero stablization potential}
\end{figure}

\begin{figure}%
\centering
\subfloat[a]{%
\label{fig:first}%
\includegraphics[height=1.5in]{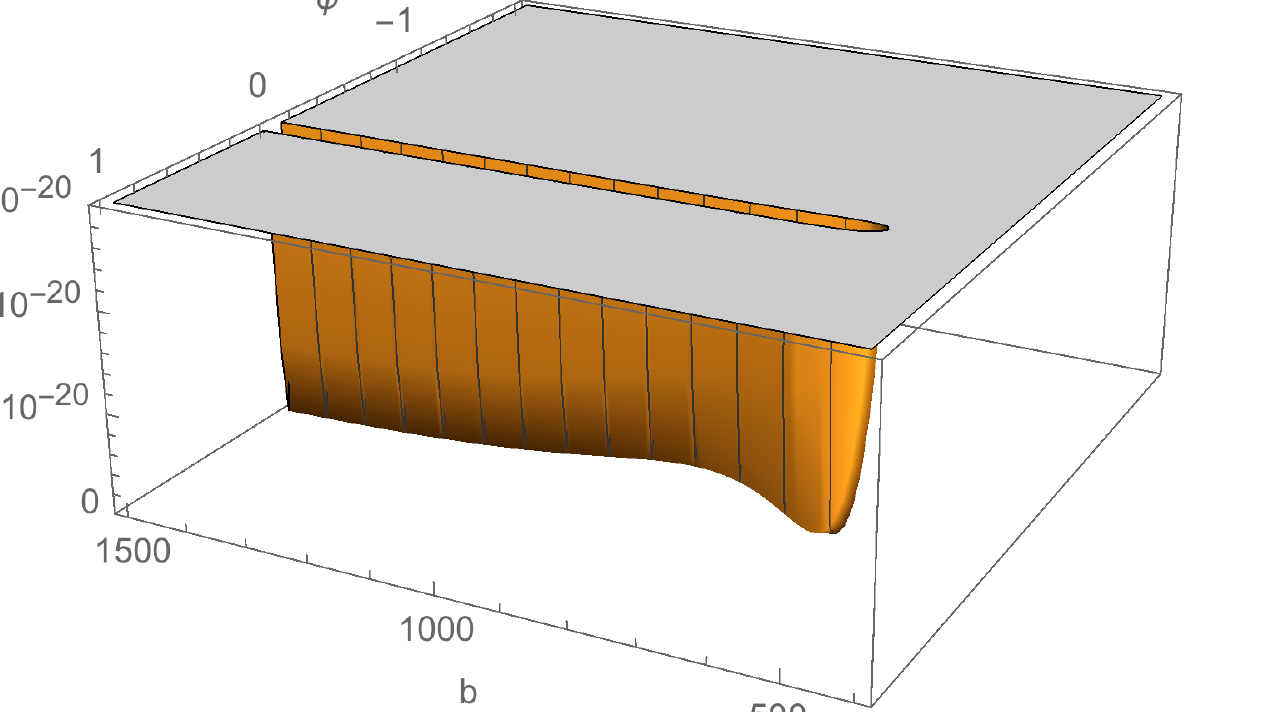}}%
\qquad
\subfloat[b]{%
\label{fig:second}%
\includegraphics[height=1.5in]{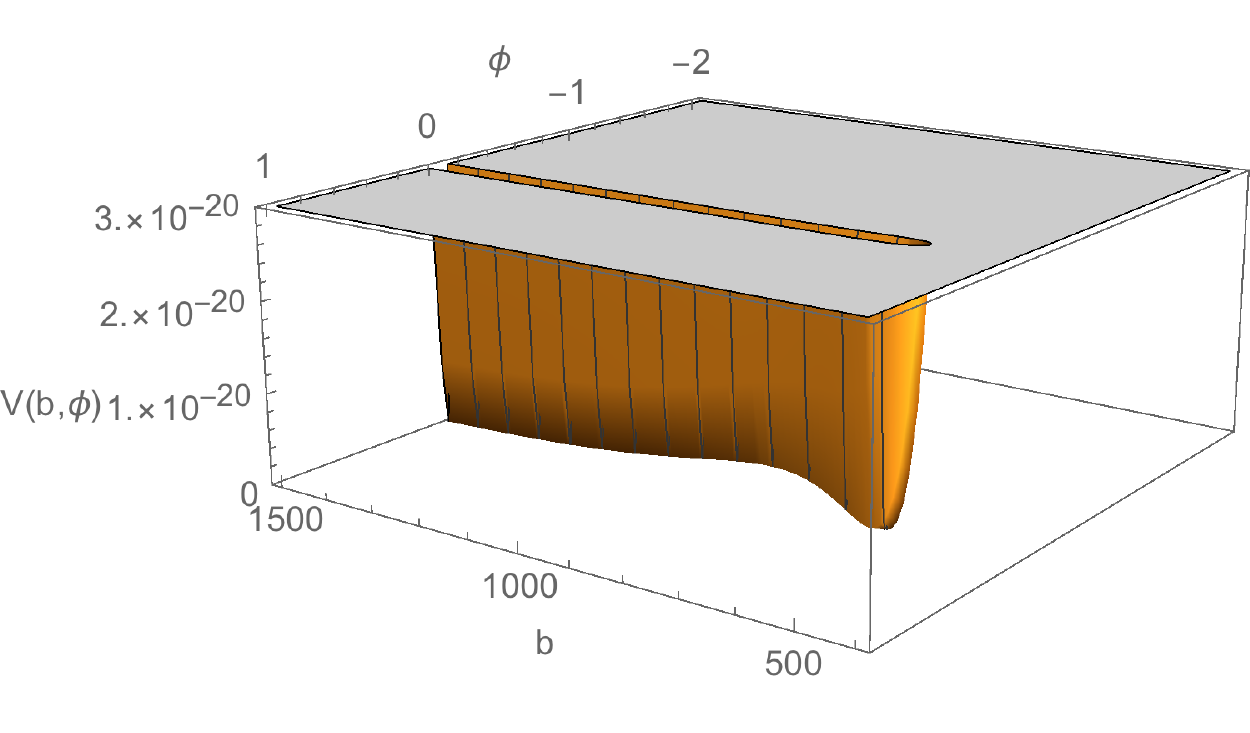}}%
\caption{[a] Dilaton-Radion potential for flux $f=1092$ and dilaton stabilization potential proportional to ${(\sinh (2\phi ))^2}$. The potential admits a local minimum with positive reduced  effective four dimensional cosmological constant.  [b] Dilaton-Radion potential for flux $f=1092$ and dilaton stabilization potential proportional to $\phi^2$. The potential admits a local minimum with positive reduced  effective four dimensional cosmological constant.The potential looks similar to the above hyperbolic sine squared stabilization potential}
\end{figure}

\section{Dark matter and bi-fundamental fermion portal field in the Compactified $SO(16) \times SO(16)'$ nonsupersymmetric model }

The $SO(16) \times SO(16)'$ non-supersymmetric string has fermions in the $(1,128_s')$ spinor representation and gauge bosons in the $(1,120_a) $ adjoint representation of $SO(16)'$. These states interact with the visible sector contained in $SO(16)$ gravitationally in a manner similar to the $E_8'$ states of the supersymmetric heterotic string \cite{Kolb:1985bf} as well as through a bi-fundamental fermion portal field. Bi-fundamental fields have also been studied in cosmology where they can lead to a network of flux tubes and cosmic strings with implications for the early Universe and galaxy formation \cite{Schellekens:2017clt}\cite{Vachaspati:2008wi}. These bi-fundamental matter fields can serve a a portal for non-gravitational interaction between the visible and hidden sectors as they are charged under both groups and can play an import role in the phenomenological consequences of the theory including dark matter production in cosmology and in accelerator experiments.

Hidden sector gauge groups smaller than $SO(16)'$ may be preferred when the Hidden sector contains dark matter candidates in the form of self interacting hidden glueballs \cite{Faraggi:2000pv}\cite{Acharya:2017szw}\cite{Boddy:2014yra}
\cite{Kribs:2016cew}\cite{Soni:2016gzf}\cite{Halverson:2016nfq}
\cite{Forestell:2016qhc}\cite{Forestell:2017wov}\cite{Batell:2009di}
\cite{Pospelov:2007mp}\cite{Juknevich:2009gg}\cite{Cline:2013zca}
\cite{Dienes:2016vei}\cite{Hambye:2009fg}\cite{Halverson:2018olu}. This follows from renormalization group analysis which relates the glueball mass scale to the reheating temperature. Compactifications which break the hidden $SO(16)'$ can in principle realize this scenario in non-supersymmetric model building using non-supersymmetric orbifolds. Also with non-supersymmetric orbifolds one can see the Higgs field emerge as a extradimensional component of the Hiiggs field. For example in \cite{Font:2002pq} A. Font found $(10,16')$ scalar multiplets in $SO(10)\times SO(16)'$ in nonsupersymmetric orbifold compactification on $T^6/Z_3$ orbifold which are in $SO(10)$ representations similar to those used in GUT Higgs models.

\begin{figure}
\hfill
\subfloat(a){\includegraphics[width=5cm]{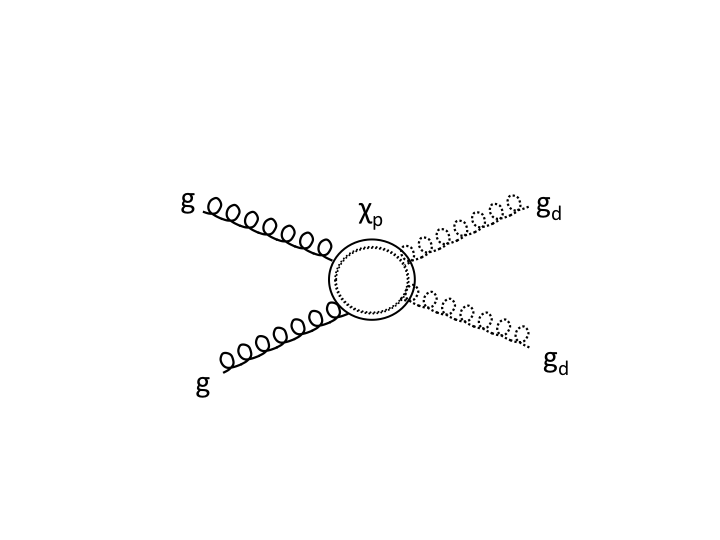}}
\hfill
\subfloat(b){\includegraphics[width=5cm]{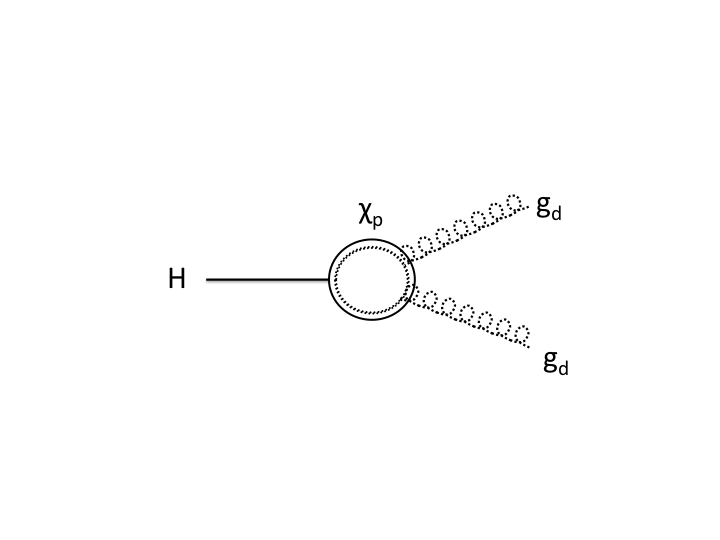}}
\hfill
\caption{(a) Dark gluon $g_d$ production through one-loop effects of a portal bi-fundamental fermion field $\chi_p$. (b) Higgs decay to Dark gluons $g_d$ through one-loop effects of a portal bi-fundamental fermion field $\chi_p$.}
\end{figure}

\subsection{Bifundamental fermion portals in accelerator experiments}

A potential scattering experiment producing two dark gluons through the interaction with a portal fermion is shown in figure 5 (a). We note that the diagram is similar to the light by light scattering diagram studied in \cite{Karplus:1950zz}\cite{Karplus:1950zza}\cite{Breit:1934zz}\cite{Euler:1936oxn} so we will use that calculation to guide an order of magnitude analysis of the amplitude in figure 5 (a). In particular the polarization amplitude $M_{++--}$ can be estimated to be \cite{Berestetsky:1982aq}\cite{Schwartz:2013pla}
\begin{equation}{M_{ +  +  -  - }} \approx  - \frac{{g_s^2g_{s'}^2({s^2} + {t^2} + {u^2})}}{{15m_{\chi}^4}}{N_{ss'}}\end{equation}
which is valid for energies far less the the portal field mass $\omega \ll m_\chi$. In this formula $g_s$ is the strong coupling, $g_{s'}$ is the dark gauge coupling, $s$, $t$, and $u$ are the kinematic Lorentz invariants,$m_p$ is the portal fermion mass and $N_{ss'}$ is a numerical factor.

The cross section is determined by integration over the square of the Matrix elements for various polararizations and is given by:
\begin{equation}\sigma  \approx N_{ss'}^2\frac{{973}}{{10125\pi }}\alpha _s^2\alpha _{s'}^2\frac{1}{{m_\chi^2}}{\left( {\frac{\omega }{{{m_\chi}}}} \right)^6}\end{equation}
where again this valid for $\omega \ll m_\chi$, $\alpha_s = g_s^2/4\pi$ and $\alpha_{s'} = g_{s'}^2/4\pi$.
For large energies $\omega  \gg {m_\chi}$ one has the expression for the cross section:
\begin{equation}\sigma  \approx N_{ss'}^24.7\alpha _s^2\alpha _{s'}^2\frac{1}{{{\omega ^2}}}\end{equation}
If the mass of the $\chi$ field is much greater than the center of mass energy of the initial gluons than the cross section falls inversely proportion to the eight power of the $\chi$ mass with a quadratic enhancement for large center of mass energy.


More refined estimates can be made using the effective field theory formalism.
For nonabelian groups we have the effective action between the  visible gluon $g$ and dark gluon $g_d$ given by:\cite{Quevillon:2018mfl}\cite{Karasik:2019bxn}
$$L_{eff}^{(8)} = {\alpha _1}\frac{{{g^2}g{'^2}}}{{6!{\pi ^2}m_\chi ^4}}F_{\mu \nu }^a{F^{a\mu \nu }}{F'}_{\mu '\nu '}^{a'}F{'^{a'\mu '\nu '}} + {\alpha _2}\frac{{{g^2}g{'^2}}}{{6!{\pi ^2}m_\chi ^4}}F_{\mu \nu }^a{{\tilde F}^{a\mu \nu }}{F'}_{\mu '\nu '}^{a'}\tilde {F'}^{a'\mu '\nu '}$$
\begin{equation} + {\alpha _3}\frac{{{g^2}g{'^2}}}{{6!{\pi ^2}m_\chi ^4}}F_{\mu \nu }^aF{'^{a'\mu \nu }}F_{\mu '\nu '}^aF{'^{a'\mu '\nu '}} + {\alpha _4}\frac{{{g^2}g{'^2}}}{{6!{\pi ^2}m_\chi ^4}}F_{\mu \nu }^a\tilde F{'^{a'\mu \nu }}F_{\mu '\nu '}^a\tilde F{'^{a'\mu '\nu '}}\end{equation}
where $F_{\mu \nu }^a$ and $F_{\mu '\nu '}^{a'}$ are the visible and dark field strengths and ${{\tilde F}^{\mu \nu a}}$ and ${{\tilde F}^{\mu '\nu 'a'}}$ are their duals.
The coefficients were determined in \cite{Quevillon:2018mfl} to be:
$${\alpha _1} = {\alpha _3}/2 = {I_2}(R){I_2}(R')$$
\begin{equation}{\alpha _2} = {\alpha _4}/2 = \frac{7}{4}{I_2}(R){I_2}(R')\end{equation}
with $I_2(R)$ defined by the lie algebra generators through $tr\left( {T_R^aT_R^b} \right) = {I_2}(R){\delta ^{ab}}$ with $I_2(R)$ normalized to $1/2$ or $1$ fro the fundamental representation for $SO(N)$ and $SU(N)$ groups respectively. Using these coefficients one can calculate the  amplitude and cross section for dark gluon production for center of mass energy below the mass of the portal fermion $m_\chi$. One can estimate fragmentation functions to dark glueballs using nonperturbative methods similar to that which is done for hadronization. These dark matter production mechanisms are actively being search for at the LHC \cite{Aaboud:2018arf}. As the cross section goes up with center of mass energy until one reaches the $\chi$ mass further energy upgrades to the accelerator should aid in the search for these dark matter production channels. Interestingly the decay channels are similar to those of an earlier model for a hidden strongly interacting sector due to Glashow  \cite{Glashow:1984su} with a signal given by missing energy.

\section{Higgs physics in the Compactified $SO(16) \times SO(16)'$ nonsupersymmetric model}

\begin{figure}
  \includegraphics[scale = .5]{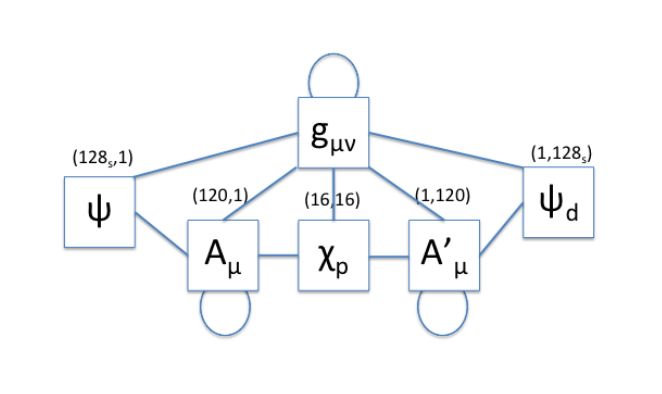}
  \caption{Fields and interactions of the $SO(16) \times SO(16)'$ nonsupersymmetric model. The hidden $SO(16)'$ sector interacts with the visible sector through the gravitational field $g_{\mu \nu}$ as well as through  a portal bi-fundamental fermion field $\chi_p$.}
  \label{}
\end{figure}

\begin{figure}%
\centering
\subfloat[a]{%
\label{fig:first}%
\includegraphics[height=1.5in]{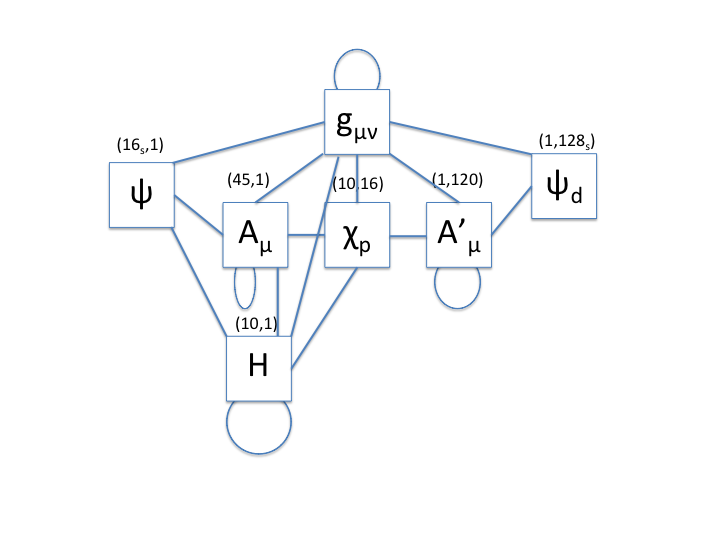}}%
\subfloat[b]{%
\label{fig:second}%
\includegraphics[height=1.5in]{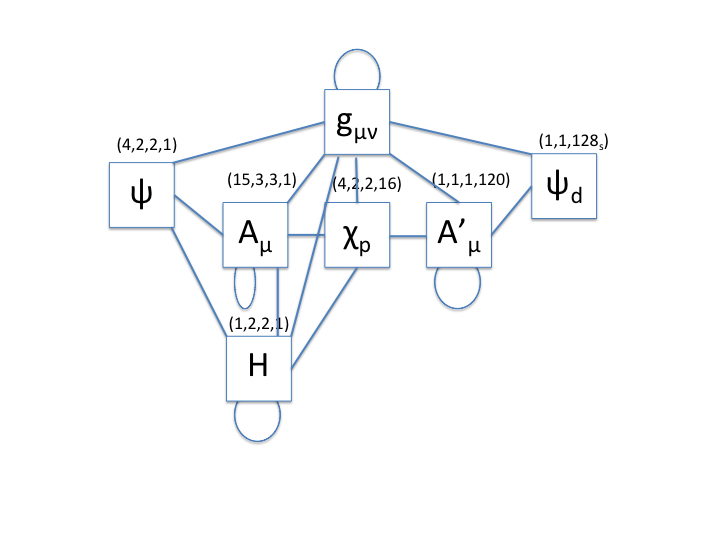}}%
\subfloat[c]{%
\label{fig:second}%
\includegraphics[height=1.5in]{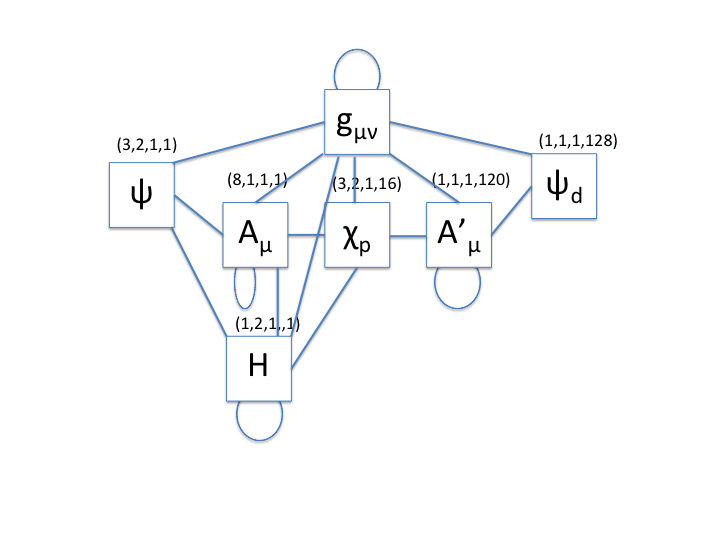}}%
\caption{[a] Fields and interactions of the compactified $SO(16) \times SO(16)'$ nonsupersymmetric model yielding $SO(10) \times SO(16)'$. The hidden $SO(16)'$ sector interacts with the visible sector (including a fundamental Higgs) through the gravitational field $g_{\mu \nu}$ as well as through  a portal bi-fundamental fermion field $\chi_p$. [b] Fields and interactions of the compactified $SO(16) \times SO(16)'$ nonsupersymmetric model yielding $SO(6) \times SO4) \times SO(16)'\cong  SU(4) \times SU(2) \times SU(2)\times SO(16)'$ which is the Pati-Salam gauge group times a hidden sector. The hidden $SO(16)'$ sector interacts with the visible sector (including a fundamental Higgs in the $(1,2,2,1')$ representation) through the gravitational field $g_{\mu \nu}$ as well as through  a portal bi-fundamental fermion field $\chi_p$ in the $(4,2,2,1')$ representation. [c] Fields and interactions of the compactified $SO(16) \times SO(16)'$ nonsupersymmetric model yielding $SU(3) \times SU(2) \times U(1) \times SO(16)'$ which is the Standard Model gauge  group times a hidden sector. The hidden $SO(16)'$   sector interacts with the visible sector (including a fundamental Higgs in the $(1,2,1,1')$ representation) through the gravitational field $g_{\mu \nu}$ as well as through  a portal bi-fundamental fermion field $\chi_p$ in the $(3,2,1,16')$ representation. Although there are no gauged scalars in the ten dimensional theory they can be be generated by extra dimensional components of the gauge field. }
\end{figure}


Although the origin of the nature of the Higgs boson is unknown it is plausible that it's origin comes from a component of higher dimensional gauge field as in Gauge-Higgs Unification. 
If so it will likely interact with the portal $\chi$ field as this field interacts with all gauge fields, both visible and hidden. In this section we will discuss the effect of this interaction with the portal field on Higgs physics.
There are at least two areas where the bifundamental portal fermion can play an important role in Higgs physics. One  is in accelerator searches for decays of the Higgs boson to dark particles.Another  way is through the effects of the portal fermion on the effective Higgs potential.

\subsection{ Higgs decay to dark matter through fermion bi-fundamental fermion in the Compactified $SO(16) \times SO(16)'$ nonsupersymmetric model }

As discussed above the   $SO(16) \times SO(16)'$ nonsupersymmetric model does not contain any gauged scalars in ten dimensions. However in the compactified theory internal components of the gauge field can play the role of the Higgs.
The Higgs is embedded in this theory by taking internal components of the $SO(16)$ gauge field. Under the decomposition $SO(16) \supset SO(10) \times SO(6)$
$$120 = (6(10),1) + (45,1) + (1,15)$$
It is the $10$ that contains the Higgs field through the Pati-Salam decomposition:
$$10 = (6,1,1) + (1,2,2)$$
Finally it is the $(1,2,2)$ representation that contains the electroweak Higgs field in the $(1,2,1)$ representation of $SU(3)\times SU(2)\times U(1)$.
As the higher dimensional gauge field is coupled to the bi-fundamental field $\chi$ it is then possible for the Higgs field to couple to $\chi$. 

For the SO(10) Grand Unified models the quarks and leptons are grouped in the spin $16$ representation
$${\psi _{16}} = ({\nu _L},{e_L},{d_{1L}},{d_{2L}},{d_{3L}},{u_{1L}},{u_{2L}},{u_{3L}},{u_{1R}},{u_{2R}},{u_{3R}},{d_{1R}},{d_{2R}},{d_{3R}},{e_R},{\nu _R})$$
The  $SO(10)$ allows baryon number violation allows for a neutrino mass. For the Pati-Salam model the quarks and leptons are in $(4,2)$ reprentations
 \begin{equation}
\psi _{(4, 2),L}=
  \begin{pmatrix}
    e_L & u_{1L} & u_{2L} & u_{3L}  \\
    \nu_L & d_{1L} & d_{2L} & d_{3L} 
  \end{pmatrix},
 \psi _{(4, 2),R}=
  \begin{pmatrix}
    e_R & u_{1R} & u_{2R} & u_{3R}  \\
    \nu_R & d_{1R} & d_{2R} & d_{3R} 
  \end{pmatrix}, 
\end{equation}
 For the Higgs decay to the dark sector the important contribution is due to the coupling of the Higgs to the portal fermion in the $(10,16')$ representation for the GUT model the $(2,2,4,16')$ repreresention for the Pati-Salam model and the $(2,3,16')$ representation for the standard model. In figure 6 and 7 we illustrate the fields and interactions for the uncompactified and compactified $SO(16) \times SO(16)'$ nonsupersymmetric heterotic model. Heterotic compactifications for $SO(10)$ GUT models, the Pati-Salam model and Standard-like models are considered in detail in \cite{Rizos:2011ib}\cite{Assel:2010wj}\cite{Faraggi:2017cnh}. Below we will consider the case of the standard model.
 
Once one has this coupling of the Higgs to the portal fermion one can consider the process of the Higgs decaying  to dark gluons as shown in figure 5 (b). The decay rate is the given by 
\cite{Gunion:1989we}\cite{Lu:2017uur}
\cite{Abe:2019wku}\cite{Cheung:2012nb}:
\begin{equation}\Gamma (H \to {g_d}{g_d}) = \frac{{\alpha _{s'}^2}}{{128{\pi ^3}}}{y^2}\frac{{m_H^3}}{{m_\chi ^2}}{f^2}\left( {\frac{{m_\chi ^2}}{{m_H^2}}} \right)\end{equation}
with an effective action between the Higgs and the dark gluons given by:
\begin{equation}L_{eff}^{(6)} = \frac{{2{\alpha _{s'}}}}{\pi }\frac{y}{{{m_\chi }}}f\left( {\frac{{m_\chi ^2}}{{m_H^2}}} \right)h(F_{\mu '\nu '}^{a'}{F^{\mu '\nu 'a'}})\end{equation}
Here $y$ is the coupling of the Higgs to the $\chi$ field and $f$ is a  function determined from the loop in  \cite{Gunion:1989we}\cite{Lu:2017uur} and plotted in figure 8.
The $f(x)$ function depends on whether the coupling of the Higgs to the $\chi$ particle is through a scalar or pseudoscalar coupling. It is is defined piece wise for a scalar coupling as:
\begin{equation}
f_S(x) =
  \begin{cases}
      4x(1 + (1 - 4x) \arcsinh \left[ \frac{1}{\sqrt {4x}}  \right] & \text{for $x < .25$} \\
      4x(1 + (1 - 4x)\arcsin \left[ \frac{1}{\sqrt {4x} } \right] & \text{for $x \ge .25$}
  \end{cases}
\end{equation}
and for a pseudoscalar coupling as:
\begin{equation}
f_{PS}(x) =
  \begin{cases}
      - 2x\arctanh \left[ \frac{1}{\sqrt {4x - 1}} \right] & \text{for $x < .25$} \\
      2x \arctan \left[ \frac{1}{\sqrt {4x - 1}} \right] & \text{for $x \ge .25$}
  \end{cases}
\end{equation}

\begin{figure}
\hfill
\subfloat(a){\includegraphics[width=5cm]{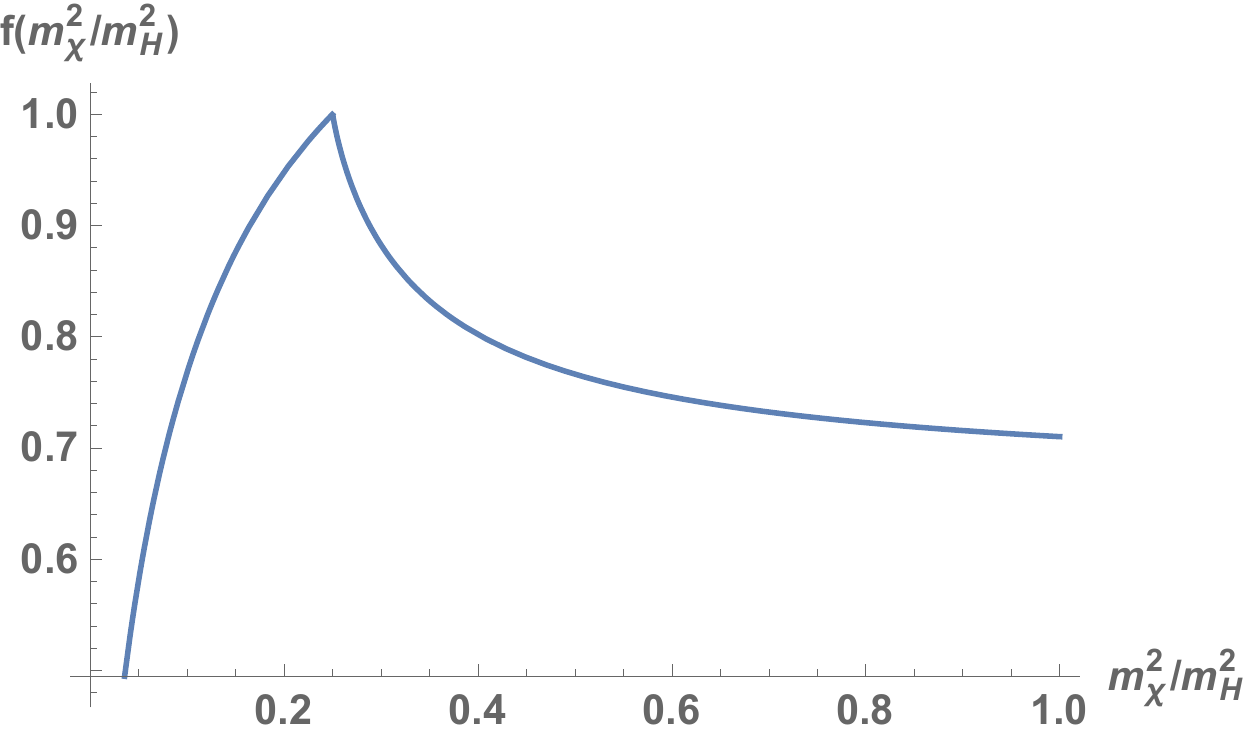}}
\hfill
\subfloat(b){\includegraphics[width=5cm]{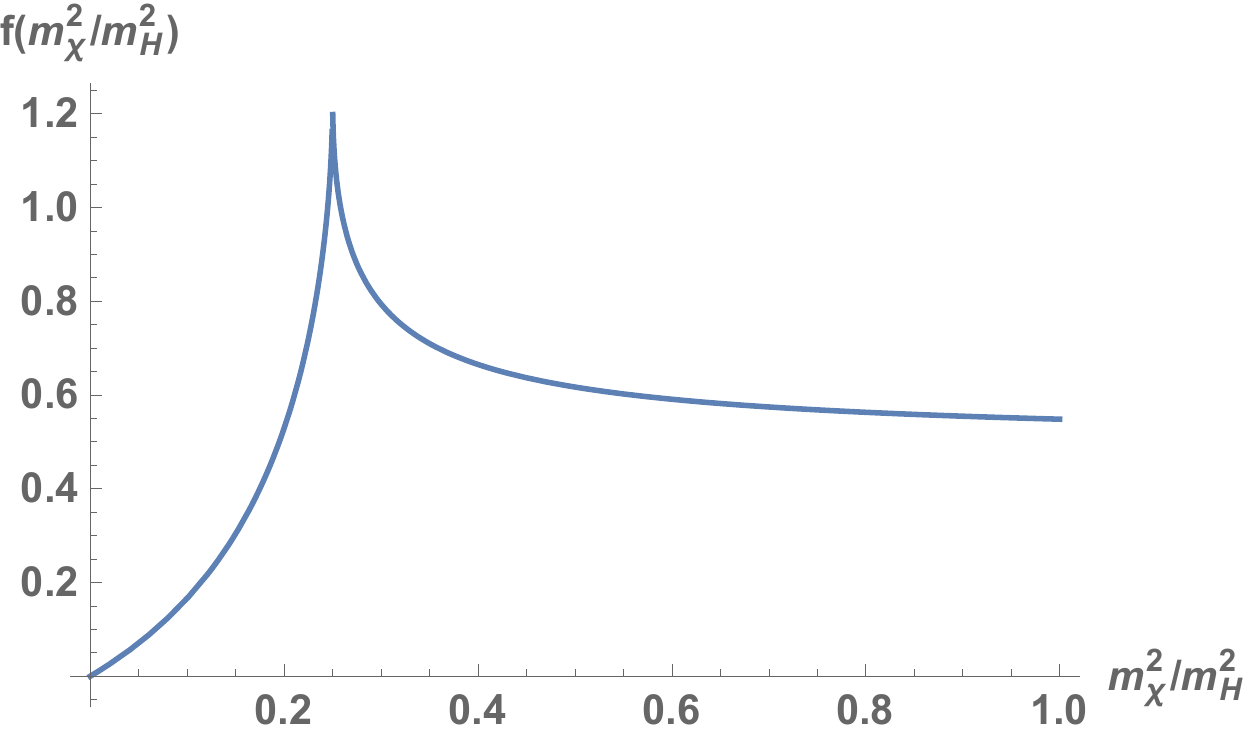}}
\hfill
\caption{$f$ function for Higgs decay to Dark gluons $g_d$ through one-loop effects of a portal bi-fundamental fermion field $\chi_p$ with (a) scalar  and (b) pseudoscalr  $\gamma_5$ coupling.}
\end{figure}

If the $\chi$ field is more massive than half the Higgs mass then the decay rate $ \Gamma (H \to {g_d}{g_d})$ is a decreasing function falling inversely proportional to the square of the $\chi$ mass. These decays to Hidden sector particles are being actively searched for at the LHC \cite{ATLAS:2017lvz}\cite{Aaboud:2018fvk}
Besides the Higgs decay to the dark gluon one can also consider Higgs decays to dark photons or dark $Z'$s\cite{Davoudiasl:2013aya}. Dark matter condidates include dark glueballs, dark baryons or dark pions. Effective actions such as above can be used to calculate the mass of these particles using Lattice methods \cite{Brower:2019oor}. If the hidden gauge group is unbroken the dark gauge group would be $SO(16)'$. Orthogonal groups are difficult to simulate although some work on $SO(16)'$ have been done in three dimensions \cite{Lau:2017aom}.The subgroup $SU(8)' \times U(1)'$ and $SU(8)'$ has been simulated in four dimensions in \cite{Lucini:2010nv} with calculations of the glueball mass. Further reductions of the hidden gauge group to $SO(6)'$ can be considered which is equivalent to $SU(4)'$ and can be efficiently simulated. The status of self interacting dark matter computations with hidden sectors is summarized in \cite{Kribs:2016cew}.

\begin{figure}
\hfill
\subfloat(a){\includegraphics[width=5cm]{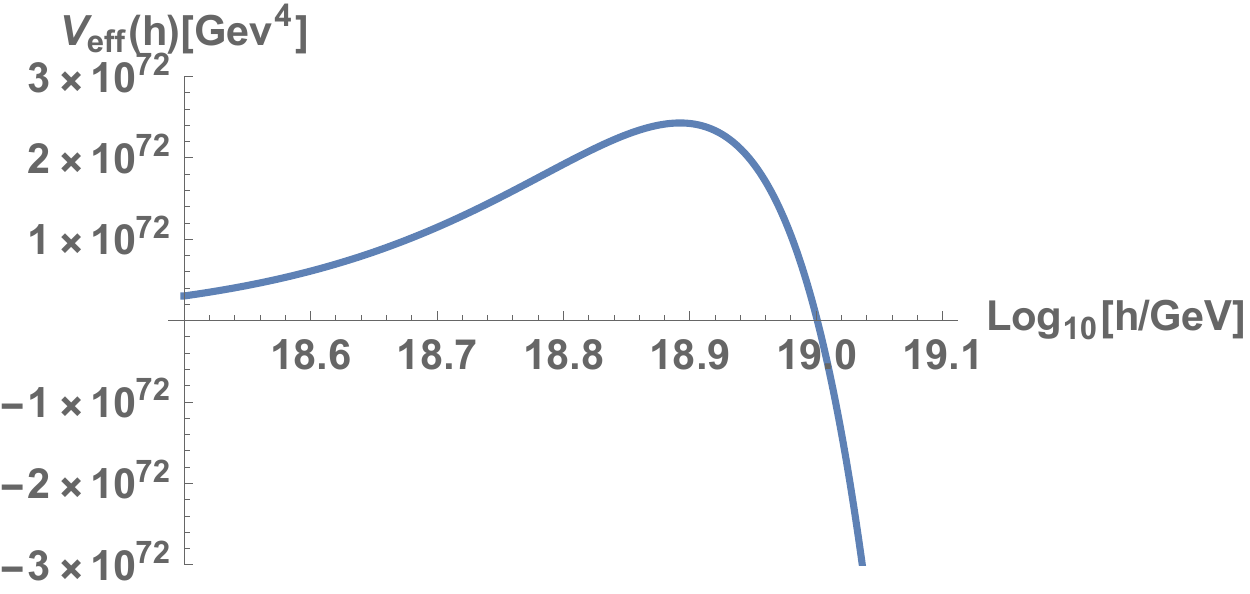}}
\hfill
\subfloat(b){\includegraphics[width=5cm]{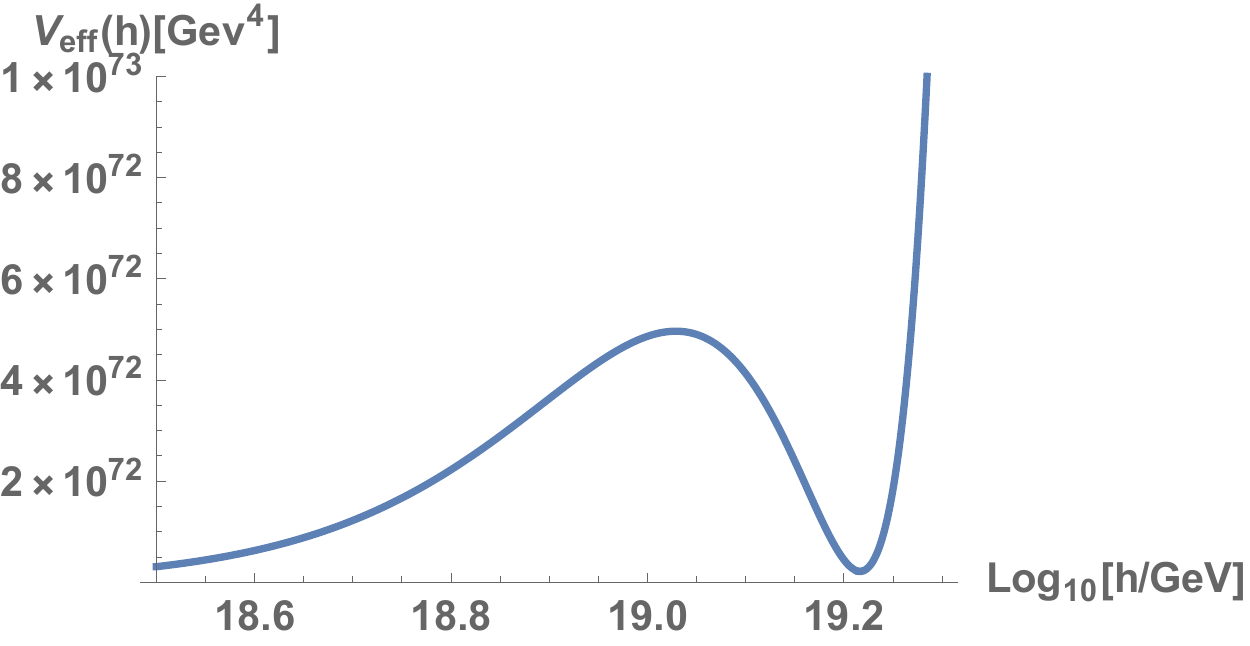}}
\hfill
\caption{(a) Effective potential for the Higgs field at large field values with the Higgs potential going through zero. (b) Effective potential for the Higgs field at large field values with the addition of a nonrenormalizable potential term $h^6$. Depending on the size of this term there can be a metastable vacuum at large values of the Higgs field.}
\end{figure}

\begin{figure}[!htb]
\minipage{0.32\textwidth}
  \includegraphics[width=\linewidth]{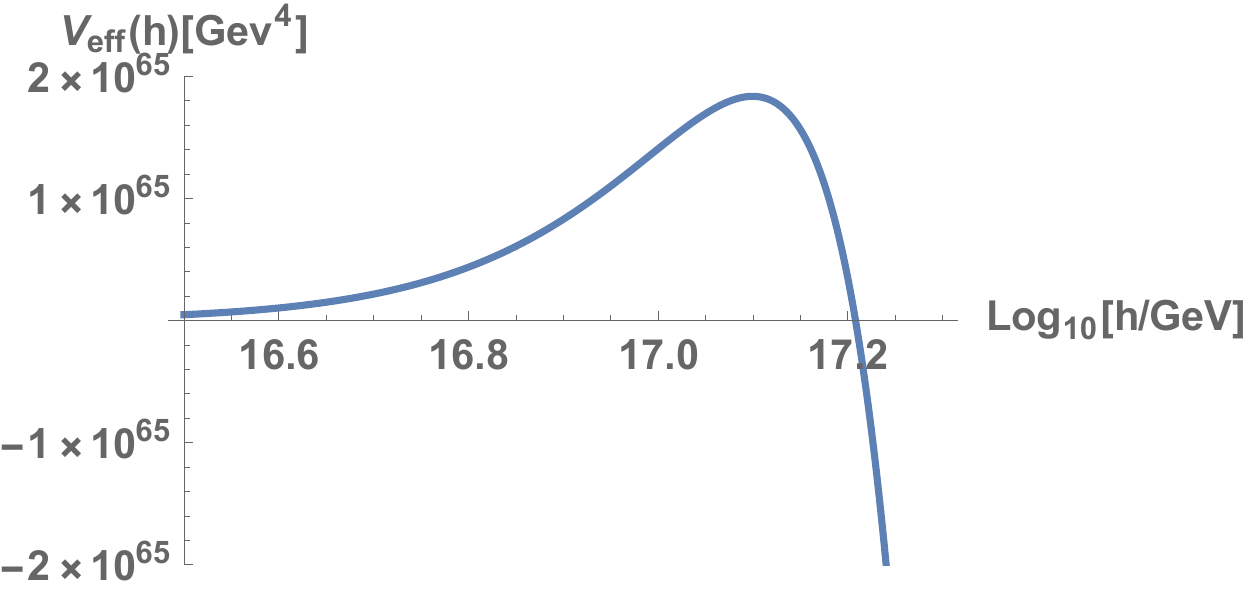}
\label{fig:awesome_image1}
\endminipage\hfill
\minipage{0.32\textwidth}
  \includegraphics[width=\linewidth]{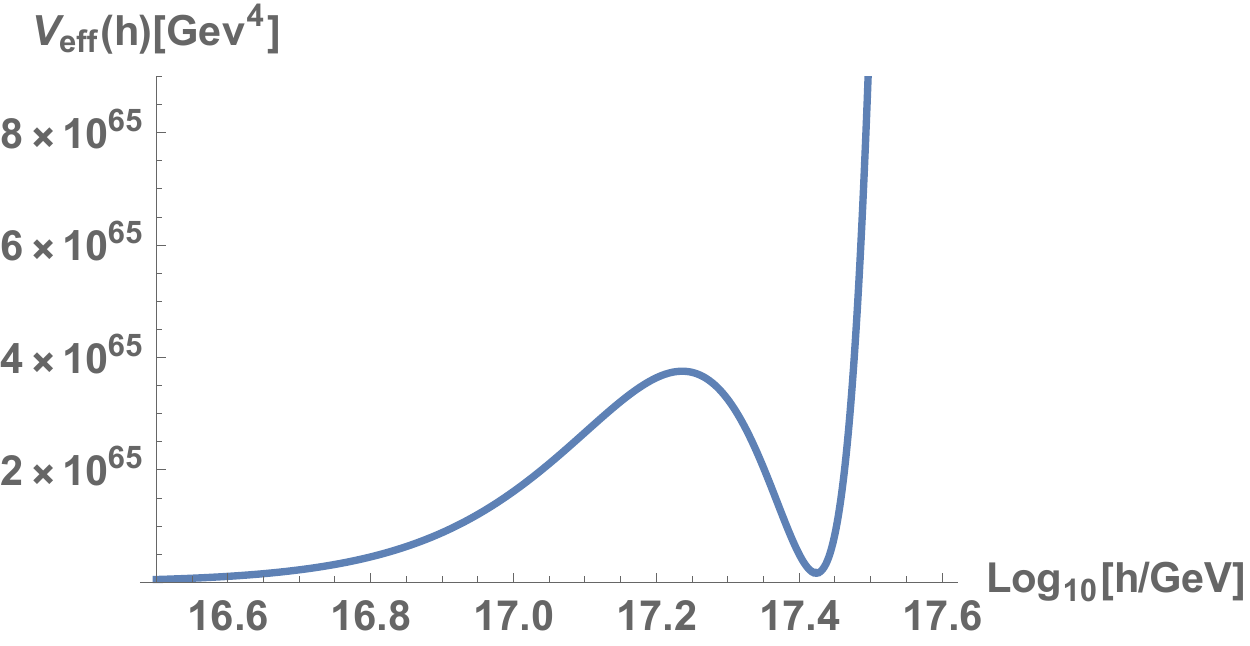}
\endminipage\hfill
\minipage{0.32\textwidth}%
  \includegraphics[width=\linewidth]{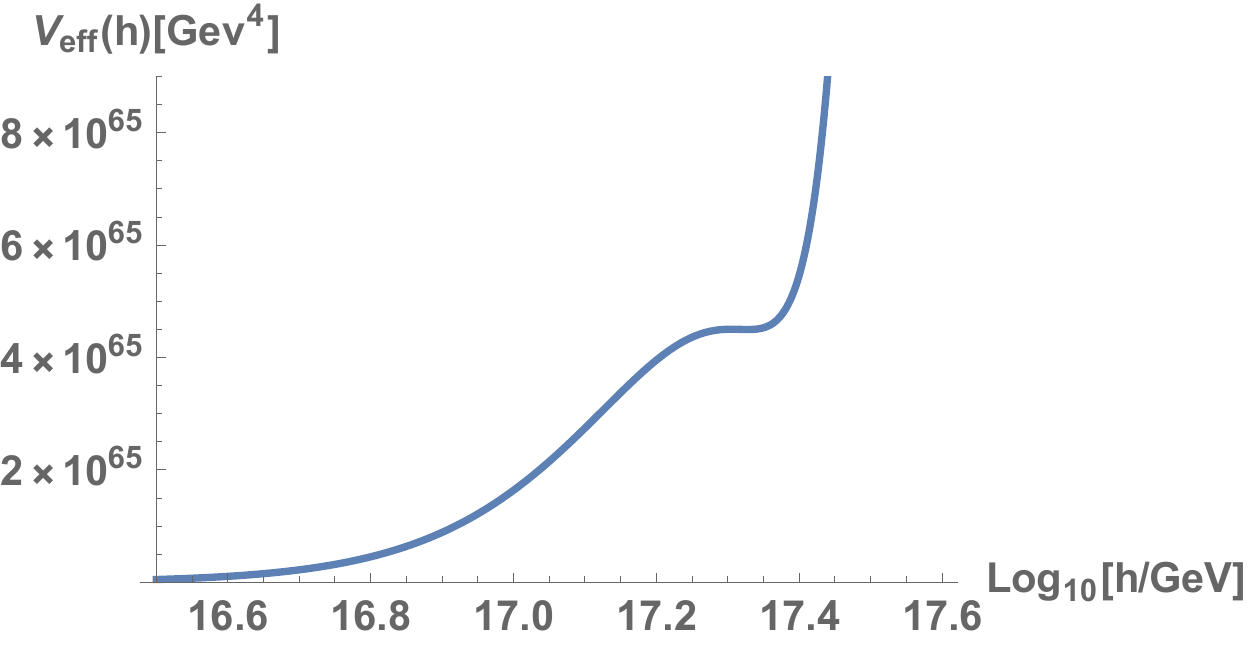}
 
\endminipage
\caption{(Left)Effective potential for the Higgs field with coupling of the Higgs to the $\chi$ field with large field values with the Higgs potential going through zero. The effect of the $\chi$ field is to lower the value of the Higgs field for which the potential goes to zero.(Middle) Effective potential for the Higgs field with coupling of the Higgs to the $\chi$ field and a nonrenormalizable term proportional to $h^6$. Depending on the size of this term there can be a metastable vacuum or (Right) hilltop potential at a lower value of the Higgs field relative to the absence of the $\chi$ field.}
\end{figure}

\subsection{Higgs potential stability bifundamental portal fermion in the $SO(16) \times SO(16)'$ nonsupersymmetric model}

The presence of the portal $ \chi$ field can effect the Higgs potential stability \cite{Gopalakrishna:2018uxn}
\cite{Held:2018cxd}\cite{Eichhorn:2015kea}\cite{Rose:2015lna}
\cite{Sondenheimer:2017jin}\cite{Bezrukov:2014ina}
\cite{Buck:2010sv}
Here we use a simple one-loop expression and $\frac{\lambda}{4} = \frac{m_h^2}{8v^2} \approx .10$ and $y_t= .93$ for illustrative purposes as was done in \cite{Schwartz:2013pla}. Using the full two loop expression and the physical values for the Higgs mass and top quark Yukawa coupling one can obtain more accurate predictions \cite{Degrassi:2012ry}. The one loop effective potential is given by:
\begin{equation}{V_{eff}}(h) = V(h) + \sum\limits_i {{{( - 1)}^{2{s_i}}}\frac{{n_d^i}}{{64{\pi ^2}}}} m_{i,eff}^4(h)\log [m_{i,eff}^2(h)/{v^2}]\end{equation}
Here the Higgs mass is related to $m$ through $m_h = \sqrt{2} m$. Where ${( - 1)^{2{s_i}}}$ is $(-1)$ for fermions and $(1)$ for bosons and $m_{i,eff} = \frac{1}{\sqrt{2}}y h$ for fermions and $m^2_{i,eff} = V''(h)$ for the scalar Higgs. Writing this out for the one-loop effective potential for the Higgs field $h$ and top quark Yukawa coupling $y_t$ we have:
\begin{equation}
{V_{eff}}(h) =  - {m^2}{h^2} + \frac{\lambda }{4}{h^4}
 + \frac{1}{{64{\pi ^2}}}{( - {m^2} + 3\lambda {h^2})^2}\log [\frac{( - {m^2} + 3\lambda {h^2})}{v^2}] - \frac{{12}}{{64{\pi ^2}}}{\left( {\frac{1}{2}y_t^2{h^2}} \right)^2}\log [\frac{y_t^2 h^2}{2 v^2}]
 \end{equation}
The behavior of the effective Higgs potential is shown in figure 9 (a). At large values of the Higgs field the potential goes through zero to large negative potential energy. The presence of nonrenormalizable terms such as those proportional to $h^6$ can cause to potential to develop a maetastable state at large values of the Higgs field as in figure 9 (b).

To take into account the effect of the $\chi$ field on the Higgs effective potential we can introduce a coupling the Higgs field through $y_\chi h$. The effective potential is then of the form:
  $$ {V_{eff}}(h) =  - {m^2}{h^2} + \frac{\lambda }{4}{h^4}  +  \frac{\lambda_6 }{6}{h^6}
    + \frac{1}{{64{\pi ^2}}}{( - {m^2} + 3\lambda {h^2})^2}\log [( - {m^2} + 3\lambda {h^2})/{v^2}]  $$
  \begin{equation}  - \frac{{4(3)}}{{64{\pi ^2}}}{\left( {\frac{1}{2}y_t^2{h^2}} \right)^2}\log [\frac{1}{2}y_t^2{h^2}/{v^2}] 
  - \frac{{4(3)(16)}}{{64{\pi ^2}}}{\left( {\frac{1}{2}y_\chi ^2{h^2}} \right)^2}\log [\frac{1}{2}y_\chi ^2{h^2}/{v^2}] \end{equation}
Note the $h_\chi$ can be small realitive to $h_t$ as the $\chi$ field field can get the majority of it's mass through interactions in the dark sector such as through a possible Yukawa coupling with a dark Higgs. In such a situation the $\chi$ field still does have an effect on the stability of the Higgs potential however. In figure 10 we see that the effect of the $\chi$ field is to move the point where the Higgs potential goes through zero to lower energies. Again introducing a $h^6$ nonrenormalizable potential \cite{Sondenheimer:2017jin} one can convert this to a metastable vacuum  at reduced values of the Higgs field compared to the usual standard model involving the top quark.

To obtain more accurate descriptions of the effect of the $\chi$ field on the Higgs potential of can use the renormalization group equations.  In the notation of \cite{Gopalakrishna:2018uxn} with $N_c'$ hidden colors to the one loop level we have:
$$\beta _\lambda ^{(1)} = \frac{1}{{16{\pi ^2}}}(24{\lambda ^2} + 4{N_c}y_t^2\lambda  - 2{N_c}y_t^4 + 4{N_c}'y_\chi ^2\lambda  - 2{N_c}'y_\chi ^4 +  \ldots )$$
$$\beta _{{y_t}}^{(1)} = \frac{1}{{16{\pi ^2}}}\left( {\frac{{(3 + 2{N_c})}}{2}y_t^3 + {N_c}'{y_t}y_\chi ^2 +  \ldots } \right)$$
\begin{equation}\beta _{{y_\chi }}^{(1)} = \frac{1}{{16{\pi ^2}}}\left( {\frac{{(3 + 2{N_c}')}}{2}y_\chi ^3 + {N_c}{y_\chi }y_t^2 +  \ldots } \right)\end{equation}
where the extra terms involve gauge boson interactions. Solving these RGEs one can obtain $\lambda(h)$ and then the effective potential is approximately given by $\frac{\lambda(h)}{4} h^4 + \frac{\lambda_6}{6} h^6$.

\section{Cosmology of the $SO(16) \times SO(16)'$ nonsupersymmetric model}

\subsection{Dark Glueball and bifundamental portal fermion  an the Cosmology of $SO(16) \times SO(16)'$ nonsupersymmetric model}

The cosmology of non-abelian hidden sectors has received a lot of interest lately 
\cite{Kribs:2016cew}\cite{Soni:2016gzf}\cite{Halverson:2016nfq}
\cite{Forestell:2016qhc}\cite{Forestell:2017wov}\cite{Batell:2009di}
\cite{Pospelov:2007mp}\cite{Juknevich:2009gg}\cite{Cline:2013zca}
\cite{Dienes:2016vei}\cite{Davoudiasl:2013aya}\cite{Hambye:2009fg}. These hidden sectors are constrained by their effects on Big Bang Nucleosythesis, the cosmic ray background and sources of cosmic and gamma rays. In \cite{Forestell:2016qhc}\cite{Forestell:2017wov} the bounds were estimated for cosmology for dark hidden sectors in the presence of connector or portal fields connecting the visible and hidden sectors. They found if one or more of the dark glueballs were stable they could potentially make up the dark matter in the Universe. For dark matter gauge groups such as $SO(N)'$ C-odd states can be  heavier than the $0^{++}$ ground state and there are less constraints on the C-odd states in these models \cite{Forestell:2016qhc}\cite{Forestell:2017wov}.

Because of the dark glueball dark matter and portal fermion connecting it to to the standard model the $SO(16) \times SO(16)'$ nonsupersymmetric model has an interesting cosmology. As a starting point consider the Hamiltonian formulation using the ansatz:
\begin{equation}ds^2 = -N^2dt^2 + a^2 d\Omega^2_3 + b^2( d\Omega^{(1)2}_2+ d\Omega^{(2)2}_2 + d\Omega^{(3)2}_2)\end{equation}
We will consider cosmologies with a radion field, dilaton field as well as dark matter gauge fields.

\subsection{Hamiltonian cosmology with the radion}

First consider the dilaton and gauge fields turned off and consider the Lagrangian with the radion field $b$.
\begin{equation}S = \frac{1}{2} (4 \pi)^3 \int {dt\{  - \frac{{6a{{\dot a}^2}}}{N}}  + 24\frac{{{{\dot b}^2}{a^3}}}{{{b^2}N}} + 6aN + {a^3}{b^{ - 6}}N(\frac{{6}}{{{b^2}}} - 2\lambda_{10}  - \frac{f^2}{{{b^{4}}}})\} \end{equation}
The analog of the Friedmann equation is:
\begin{equation}
 - \frac{3a\dot a^2}{N}  + 12\frac{{{{\dot b}^2}{a^3}}}{b^2N} - 3a N + {a^3}{b^{ - 6}}N(-\frac{3}{b^2} + \lambda_{10}  + \frac{f^2}{2b^4})=0
 \end{equation}
The canonical momentum are:
$${\pi _a} =  - 6\frac{{a\dot a}}{N}(4 \pi)^3 $$
\begin{equation}{\pi _b} = 24\frac{{\dot b{a^3}}}{{{b^2}N}}(4 \pi)^3 \end{equation}
and the Hamiltonian constraint is:
\begin{equation}H =  (4 \pi)^{-6} ( - \frac{{\pi _a^2N}}{{12a}} + \frac{{\pi _b^2{b^2}N}}{{48{a^3}}}) - 3aN + {a^3}{b^{ - 6}}N( - \frac{{3}}{{{b^2}}} + \lambda_{10}  + \frac{f^2}{{{2 b^{4}}}}) = 0\end{equation}
Then one finds the effective potential for the radion $b$ used in section 3 to be:

\begin{equation}V_{eff}(b)= (4 \pi)^3 {a^3}{b^{ - 6}}N( - \frac{{3}}{{{b^2}}} + \lambda_{10}  + \frac{f^2}{{{2 b^{4}}}})\end{equation}
Defining $\bar b = 1/b$  we have:
\begin{equation}
H =  (4 \pi)^{-6} ( - \frac{{\pi _a^2N}}{{12a}} + \frac{ {\pi _{\bar b}}^2{\bar b}^2 N }{48 a^3}) - 3aN + {a^3}{ {\bar b}^{  6}}N( - 3 {\bar b}^2 + \lambda_{10}  + \frac{1}{2} f^2 {\bar b}^4) = 0\end{equation}
This form can be useful if one quantizes the constraint and promotes the momentum and fields to operators to form a quantum cosmology. In particular the polynomial form of the potential is more straightforward to realize quantum cosmology on early versions of quantum computers 
\cite{Kim:2019opp}\cite{Kocher:2018ilr}. We will further discuss the quantum cosmology of the model in section 9.

\subsection{Hamiltonian cosmology with the dilaton}
Finally adding a dilaton field to the above one has the action:
\begin{equation}S =  \frac{1}{2} (4 \pi)^3 \int {dt\{  - \frac{{6a{{\dot a}^2}}}{N}}  + 24\frac{{{{\dot b}^2}{a^3}}}{{{b^2}N}} + 2\frac{{{{\dot \phi }^2}{a^3}}}{N} - 12\frac{{\dot \phi \dot b{a^3}}}{{bN}} + 6aN + {a^3}{e^{2\phi }}{b^{ - 6}}N(\frac{{6}}{{{b^2}}} - 2\lambda {e^{2\phi }} - \frac{f^2}{{{b^{4}}}})\} \end{equation}

The analog of the Friedmann equation is then:
\begin{equation}
- \frac{3a{\dot a}^2}{N}  + 12\frac{{{{\dot b}^2}{a^3}}}{{{b^2}N}} + \frac{{{{\dot \phi }^2}{a^3}}}{N} - 6\frac{{\dot \phi \dot b{a^3}}}{{bN}} - 3aN + {a^3}{e^{2\phi }}{b^{ - 6}}N(-\frac{{3}}{{{b^2}}} + \lambda_{10} {e^{2\phi }} + \frac{f^2}{2 b^{4}}) = 0
\end{equation}
with canonical momentum:
$${\pi _a} =  - 6\frac{{a\dot a}}{N}(4 \pi)^3$$
$$b{\pi _b} = (24\frac{{\dot b{a^3}}}{{bN}} - 6\frac{{\dot \phi {a^3}}}{N})(4 \pi)^3$$
\begin{equation}{\pi _\phi } = (2\frac{{\dot \phi {a^3}}}{N} - 6\frac{{\dot b{a^3}}}{{bN}})(4 \pi)^3\end{equation}
and Hamiltonian constraint:
\begin{equation}H =  (4 \pi)^{-6} (- \frac{{\pi _a^2N}}{{12a}} + \frac{{\pi _b^2{b^2}N}}{{12{a^3}}} + \frac{{\pi _\phi ^2N}}{{{a^3}}} + \frac{{{\pi _\phi }{\pi _b}bN}}{{2{a^3}}}) - 3aN + {a^3}{e^{2\phi }}{b^{ - 6}}N( - \frac{{3}}{{{b^2}}} + \lambda_{10} {e^{2\phi }} + \frac{f^2}{{{2 b^{4}}}}) = 0\end{equation}
In this form one can identify the effective potential studied in section 3 form
\begin{equation} V_{eff}(b,\phi) = (4 \pi)^3{e^{2\phi }}{b^{ - 6}}( - \frac{{3}}{{{b^2}}} + \lambda_{10} {e^{2\phi }} + \frac{f^2}{{{2 b^{4}}}})\end{equation}
Finally an interesting extension is to consider Higgs-dilaton-radion cosmology given by combining the Higgs potential from section 4 with the above potential:
\begin{equation}V_{eff}(h,b,\phi) = V_{higgs}(h) + (4 \pi)^3{e^{2\phi }}{b^{ - 6}}( - \frac{{3}}{{{b^2}}} + \lambda_{10} {e^{2\phi }} + \frac{f^2}{{{2 b^{4}}}})\end{equation}
similar to the Higgs-dilaton cosmology considered in \cite{GarciaBellido:2011de}\cite{GarciaBellido:2012zu}
\cite{Bezrukov:2012hx}\cite{Rubio:2014wta}\cite{Trashorras:2016azl}
\cite{Casas:2017wjh}\cite{Sloan:2018osd}.

\subsection{Cosmology with Dark matter gauge fields}

In this subsection we derive the Friedman equation for the $SO(16) \times SO(16)'$ nonsupersymmetric model with gauge fields with the $\chi$ field integrated out so that the constraint only involves bosonic fields. For simplicity we first study the visible gauge group $SU(2)$ and hidden gauge group $SU(2)'$. Then we will generalize to the larger visible and hidden gauge groups. For the interaction between matter and dark matter we restrict ourselves to the second term in so that we may use an analysis similar to \cite{Maleknejad:2011jw}\cite{Maleknejad:2012fw} who studied gauge-flation. In this analysis one considers an action:
\begin{equation}
S = \int {{d^4}x\sqrt { - g}  {\frac{R}{{16\pi G}} - \frac{1}{4}F_{\mu \nu }^a F_a^{\mu \nu }- \frac{1}{4}{F'}_{\mu \nu }^a {F'}_a^{\mu \nu } + \frac{\kappa }{{384}}{({\varepsilon ^{\mu \nu \lambda \sigma }}F_{\mu \nu }^a F_{\lambda \sigma }^a})({\varepsilon ^{\mu' \nu' \lambda' \sigma' }}{F'}_{\mu' \nu' }^{a'}{F'}_{\lambda' \sigma' }^{a'})}}\\
\end{equation}
where we have defined:
\begin{equation}\frac{3}{2}\frac{\kappa }{{{{(4!)}^2}}} = \frac{{14}}{{45}}\frac{{{\alpha _s}{\alpha _s}'}}{{m_\chi ^4}}\end{equation}
with the ansatz:
\[A_i^a = \phi_v (t)\delta _i^a,            {\rm A}_0^a = 0\]
\begin{equation}{A'}_i^{a'} = \phi_d (t)\delta _i^{a'},            {\rm A'}_0^{a'} = 0\end{equation}
where $\phi_u$ and $\phi_d$ refer to the components of the visible and hidden gauge potential respectively. The action reduces to:
\begin{equation}
L = \frac{3}{2}(\frac{{{{\dot \phi_v }^2}}}{{{a^2}}} - {g^2}\frac{{{\phi_v ^4}}}{{{a^4}}} + \frac{{{{\dot \phi_d }^2}}}{{{a^2}}} - {g'^2}\frac{{{\phi_d ^4}}}{{{a^4}}} + \kappa g g'\frac{{{{\dot \phi }_v}\phi _v^2{{\dot \phi }_d}\phi _d^2}}{{{a^6}}})
\end{equation}
The energy density and pressure are then:
\begin{equation}\rho  = {\rho _{YM}} + {\rho _\kappa },                P = \frac{1}{3}{\rho _{YM}} - {\rho _\kappa }\end{equation}
where
\begin{equation}{\rho _{YM}} = \frac{3}{2}(\frac{{{{\dot \phi_v }^2}}}{{{a^2}}} + {g^2}\frac{{{\phi_v ^4}}}{{{a^4}}}) +\frac{3}{2}(\frac{{{{\dot \phi_d }^2}}}{{{a^2}}} + {g'^2}\frac{{{\phi_d ^4}}}{{{a^4}}}),               {\rho _\kappa } = \frac{3}{2}\kappa g g'\frac{{{{\dot \phi }_v}\phi _v^2{{\dot \phi }_d}\phi _d^2}}{{{a^6}}}
\end{equation}
and then Einstein's equations become:
\[
\frac{{{{\dot a}^2}}}{{{a^2}}} = \frac{1}{2}\left( {\frac{{{{\dot \phi_v }^2}}}{{{a^2}}} + {g^2}\frac{{{\phi_v ^4}}}{{{a^4}}} + {\frac{{{{\dot \phi_d }^2}}}{a^2}}} + {g'^2}\frac{{{\phi_d ^4}}}{{{a^4}}} + \kappa g g'\frac{{{{\dot \phi }_v}\phi _v^2{{\dot \phi }_d}\phi _d^2}}{a^6} \right)\\
\]
\begin{equation}
\frac{{\ddot a}}{a} - \frac{{{{\dot a}^2}}}{{{a^2}}} =  - \left( {\frac{{{{\dot \phi_v }^2}}}{{{a^2}}} + {g^2}\frac{{{\phi_v ^4}}}{{{a^4}}}} +{\frac{{{{\dot \phi_d }^2}}}{{{a^2}}} + {g'^2}\frac{{{\phi_d ^4}}}{{{a^4}}}} \right)
\end{equation}
One can generalize this hidden gauge sector cosmology this to arbitrary gauge groups. For example $SO(10) \times SO(16)'$ which is large enough to contain the standard model as well as a hidden gauged dark sector. Following \cite{Bertolami:1990je} \cite{Rudolph:1997gz}\cite{Rudolph:1997se}\cite{Cavaglia:1996ek} we write  the ansatz for the metric as:
\begin{equation}d{s^2} =  - {N^2}(t)d{t^2} + {a^2}(t){\omega ^i} \otimes {\omega ^i}\end{equation}
where $\omega^i$ are the three Maurer-Cartan forms satisfying 
$ d{\omega ^i} = {\varepsilon ^{ijk}}{\omega ^j} \wedge {\omega ^k}$.
For $SO(N) \times SO(N')$ one uses the ansatz:
$$A(t) = {A_0}(t)dt + {A_i}(t){\omega ^i}$$
\begin{equation}A'(t) = {A'_0}(t)dt + {A'_i}(t){\omega ^i}\end{equation}
and the definitions
  $$ {A_0}(t) = \frac{1}{2}{\Lambda ^{IJ}}(t){T_{IJ}}  $$ 
  $$ {A_i}(t) = \left[ {1 + {\phi _v}(t)} \right]{T_i} + \phi _v^I(t){T_{iI}}  $$ 
  $$ A{'_0}(t) = \frac{1}{2}{\Lambda ^{I'J}}'(t){T'_{I'J'}}  $$
  \begin{equation} A{'_i}(t) = \left[ {1 + {\phi _d}(t)} \right]{T_i} + \phi _d^{I'}(t){T'_{iI'}}  \end{equation}
    Here $i=1,2,3 $, and we have  $I,J=1,2,\ldots N-3 $ and $I',J' = 1,2, N'-3$. $T_i$ genarate the $SO(3)$ Lie Algebra, $T_{IJ}$, $T_{iJ}$ and  $T'_{I'J'}$, $T'_{iJ'}$  generate the $SO(N)$ and $SO(N')$ Lie algebras. The field strengths are then:
$${F_{0i}} = {{\dot \phi }_d}{T_i} + 2\dot \phi _d^I{T_{iI}} - 2\phi _d^I{\Lambda ^{IJ}}{T_{iJ}}$$
$${F_{ij}} = (1 - \phi _d^2 - \phi _d^I\phi _d^I){\varepsilon _{ikj}}{T_k} - 2{\phi _d}{\varepsilon _{ikj}}\phi _d^I{T_{kI}}$$
$${{F'}_{0i}} = {{\dot \phi }_d}{T_i} + 2\dot \phi _d^{I'}{{T'}_{iI'}} - 2\phi _d^{I'}{{\Lambda'} ^{I'J'}}{{T'}_{iJ'}}$$
\begin{equation}{F'_{ij}} = (1 - \phi _d^2 - \phi _d^{I'}\phi _d^{I'}){\varepsilon _{ikj}}{T_k} - 2{\phi _d}{\varepsilon _{ikj}}\phi _d^{I'}{T'_{kI'}}\end{equation}
    $N(t)$, $\Lambda^{IJ}$ and ${\Lambda'}^{I'J'}$ are Lagrange multiplies that impose the Hamiltonian and gauge constraints. The visible and hidden gauge fields are described by the $(N-2)$ variables $(\phi_v, \phi_v^I)$ and $(N'-2)$ variables $(\phi_d, \phi_d^{I'})$ respectively. The energy of the visible and dark  sector is then given by:
$${H_v} = \frac{3}{{2g^2}}\left\{ {\frac{a}{N}(\dot \phi _v^2 + \dot \phi _v^I\dot \phi _v^I) + \frac{N}{a}\left( {{{(1 - \phi _v^2 - \phi _v^I\phi _v^I)}^2} + 4\phi _v^2\phi _v^I\phi _v^I} \right)} \right\}$$
\begin{equation}{H_d} = \frac{3}{{2g'^2}}\left\{ {\frac{a}{N}(\dot \phi _d^2 + \dot \phi _d^{I'}\dot \phi _d^{I'}) + \frac{N}{a}\left( {{{(1 - \phi _d^2 - \phi _d^{I'}\phi _d^{I'})}^2} + 4\phi _d^2\phi _d^{I'}\phi _d^{I'}} \right)} \right\}\end{equation}
Again using the mixing term coupling to a fermionic a portal field we have:
\begin{equation}{H_{mix}} = \frac{3}{2}\kappa gg'\left[ {{\phi _v}(\dot \phi _v^I\phi _v^I) + {{\dot \phi }_v}(1 - \phi _v^2 - \phi _v^I\phi _v^I)} \right]\left[ {{\phi _d}(\dot \phi _d^{I'}\phi _d^{I'}) + {{\dot \phi }_d}(1 - \phi _d^2 - \phi _d^{I'}\phi _d^{I'})} \right]\end{equation}
varying the action with respect to $\Lambda^{IJ}$ and $\Lambda'^{I'J'}$ which have no time derivatives terms yield the gauge constraints:
$$\phi _v^I\dot  \phi _v^J - \dot  \phi _v^I\phi _v^J = 0$$
\begin{equation}\phi_d^{I'}{\dot  \phi }_d^{J'} - {\dot  \phi }_d^{I'}\phi _d^{J'} = 0\end{equation}
One can include fermion zero modes in the Hamiltonian as well.
the fermionic portal field has a zero mode mixing term between the dark and visible sector given by:
$${H_{portal}} = \frac{3}{2}a^2 N g{{\bar \chi }_{aa'}}\left[ {(1 + {\phi _v}){{\left( {{T_i}} \right)}_{ab}}{\gamma ^i} + \phi _v^I{{\left( {{T_{iI}}} \right)}_{ab}}{\gamma ^i}} \right]{\chi _{ba'}} $$
\begin{equation}+ \frac{3}{2}a^2 N g'{{\bar \chi }_{aa'}}\left[ {(1 + {\phi _d}){{\left( {T{'_i}} \right)}_{a'b'}}{\gamma ^i} + \phi _d^{I'}{{\left( {T{'_{iI'}}} \right)}_{a'b'}}{\gamma ^i}} \right]{\chi _{ab'}}\end{equation}
The total Hamiltonian constraint which yields the Friedmann equation for the system is given by:
\begin{equation} H_{grav} + H_v+ H_d + H_{mix} + H_{portal} = 0\end{equation}
where
\begin{equation}{H_{grav}} =  - \frac{3}{N}a{{\dot a}^2} - 3Na + N{\lambda _4}{a^3}\end{equation}
is the usual gravitational contribution with a four dimensional cosmological constant. 
Finally  Lattice computations can be performed to determine  from the gauge theory the equation of state, energy density and pressure \cite{Cheng:2007jq}\cite{Lau:2015cna}. These can be used to couple to the Einstein equations to derive a cosmology associated with the visible and hidden sectors similar to the treatment of QCD cosmology in \cite{McGuigan:2008pz}.
Lattice computations could be used as a model of the interaction between the visible and hidden sectors  to estimate the magnitude of the mixing component in the equation of state.

\section{Fischler-Susskind mechanism in the compactified $SO(16) \times SO(16)'$ nonsupersymmetric model}

The Fischler-Susskind mechanism is a  a way of stabilizing a nonsupersymmetric theory by cancelling one-loop tadpole amplitudes against tree level amplitudes calculated in a shifted non-Ricci flat background. Fischler and Susskind formulated their mechanism in two ways, one way using light-cone string field theory and another way using loop corrected beta functions of the two dimensional sigma model \cite{Fischler:1986ci}\cite{Fischler:1986tb}
  However one must keep in mind Limitations on the applicability of teh mechanism to nonsupersymmetric models in general  discussed in \cite{Banks:1999tr}. In particular the lack BPS states may hinder it's realization as a M(atrix) theory. In this section we discuss the Fischler-Susskind mechanism applied to the compactified $SO(16) \times SO(16)'$ nonsupersymmetric model. The Fischler-Susskind mechanism in a perturbative setting has been generalized to arbitrary loops for the $SO(16) \times SO(16)'$ nonsupersymmetric model by La and Nelson \cite{La:1989xk}\cite{La:1989kw}. 

The heterotic sigma model can be written as \cite{Hamada:1987ph}\cite{Foakes:1988wy}
\cite{Cai:1986sa}\cite{Ellwanger:1988cc}\cite{Callan:1986bc}
$$S = \frac{1}{{4\pi \alpha '}}\int {{d^2}\sigma } \{ {g_{\mu \nu }}(X){\partial _a}{X^\mu }{\partial ^a}{X^\nu } + i{{\bar \xi }^I}({\gamma ^a}{\partial _a}{\xi ^I}) + i{\xi ^I}({\gamma ^a}\omega _\mu ^{IJ}{\partial _a}{X^\mu }){\xi ^J} + \frac{1}{6}{R_{\mu \nu \sigma \lambda }}{{\bar \xi }^\mu }{\xi ^\sigma }{{\bar \xi }^\nu }{\xi ^\lambda }$$
\begin{equation}i{\lambda ^A}{\gamma ^a}{\partial _a}{\lambda ^A} - {\lambda ^A}{\gamma ^a}A_\mu ^{AB}{\partial _a}{X^\mu }{\lambda ^B} - \frac{1}{4}iF_{\mu \nu }^{AB}{{\bar \xi }^\mu }{\gamma ^a}{\xi ^\nu }{\lambda ^A}{\gamma _a}{\lambda ^B}\}\end{equation}
String Loop corrected beta functions for the heterotic sigma model involves cancellation between tree and one loop terms. The tree level term  for the genus $g=0$ contribution is given by:
\begin{equation}\beta _{\mu \nu }^{(g = 0)} = {R_{\mu \nu }} - \frac{1}{2}\alpha 'F_{\mu \sigma }^aF_\nu ^{a\sigma }\end{equation}
For this section we will use $a$ to present the adjoint indices for the product group $SO(16) \times SO(16)'$. For the string one-loop term for the genus $g=1$ contribution to the beta functions.:
\begin{equation}\beta _{\mu \nu }^{(g = 1)} = {\lambda _{10}}{g_{\mu \nu }}\end{equation}

To evoke the Fischler-Susskind mechanism one combines these two terms and works  with the total beta function contributions  as follows:
\begin{equation}\beta _{\mu \nu }^{FS} =\beta _{\mu \nu }^{(g = 0)} +  \alpha'^4\beta _{\mu \nu }^{(g = 1)} = {R_{\mu \nu }} - \frac{1}{2}\alpha 'F_{\mu \sigma }^aF_\nu ^{a\sigma }+ \alpha'^4{\lambda _{10}}{g_{\mu \nu }}\end{equation}
Demanding that this equals zero one obtains the loop corrected equation of motion with the contribution from the cosmological constant. 
Adding these contributions together and requiring that they cancel so the heterotic sigma model is conformally invariant yield the same equations (or linear combinations of them) as the loop corrected effective actions that we considered in section 3, Thus the application  of the Fischler Susskind mechanism for the $SO(16) \times SO(16)'$ nonsupersymmetric model allows one to define a consistent string model  yielding the loop corrected effective action.  The final result is that the background is shifted to a non-Ricci flat metric whose curvature depends on the one-loop value of the cosmological constant $\lambda_{10}$ whose computation was reviewed in section 3.

One can also add the dilaton and antisymmetric tensor field to these equations as in \cite{Callan:1986bc}. Indeed this is necessary for the consistency of the equations. The beta function for the dilaton and gravitational field are given by:
$${\beta ^\phi } = \frac{1}{2}R - {\nabla ^2}\phi  - \frac{1}{2}{(\nabla \phi )^2}$$
\begin{equation}{\beta _{\mu \nu }} = {R_{\mu \nu }} + 2{\nabla _\mu }{\nabla _\nu }\phi  - {\lambda _{10}}{e^{2\phi }}{g_{\mu \nu }}\end{equation}
Forming the combination:
\begin{equation}{\beta _{\mu \nu }} - {\beta ^\phi }{g_{\mu \nu }} = {T_{\mu \nu }}\end{equation}
where $T_{\mu \nu}$ is the matter stress energy condition and setting the dliaton to zero assuming one introduces a nonperturbative dilaton potential to give the dilaton a mass as in section 3 we obtain:
\begin{equation}{R_{\mu \nu }} - \frac{1}{2}R{g_{\mu \nu }} + {\lambda _{10}}{g_{\mu \nu }}= {T_{\mu \nu }}\end{equation}
which we recognize as Einstein's equations with cosmological constant with units chosen such that $8 \pi G = 1$.
We can rewrite these equations as:
\begin{equation}{R_{\mu \nu }} =\frac{1}{8}( 8{T_{\mu \nu }}- T{g_{\mu \nu }} + 2{\lambda _{10}}{g_{\mu \nu }})\end{equation}
with $T$ the trace of the matter stress energy tensor.
For the ansatz:
\begin{equation}d{s^2} =  - d{t^2} + {a^2}(t)(dx^2+dy^2+dz^2) + b_1^2(t)d\Omega _2^{(1)2} + b_2^2(t)d\Omega _2^{(2)2} + b_3^2(t)d\Omega _2^{(3)2}\end{equation}
and the ansatz for separate fluxes $f_i$ through the spheres as in section 3. The equations become \cite{Kolb:1986nj}:
\begin{equation}\begin{array}{l}
3\frac{{\ddot a}}{a} + 2(\frac{{{{\ddot b}_1}}}{{{b_1}}} + \frac{{{{\ddot b}_2}}}{{{b_2}}} + \frac{{{{\ddot b}_3}}}{{{b_3}}}) =  - \frac{1}{8}(\frac{{f_1^2}}{{b_1^4}} + \frac{{f_2^2}}{{b_2^4}} + \frac{{f_3^2}}{{b_3^4}} - 2{\lambda_{10} })\\
\frac{{\ddot a}}{a} + 2\frac{{{{\dot a}^2}}}{{{a^2}}} + 2\frac{{\dot a}}{a}(\frac{{{{\dot b}_1}}}{{{b_1}}} + \frac{{{{\dot b}_2}}}{{{b_2}}} + \frac{{{{\dot b}_3}}}{{{b_3}}}) = \frac{1}{8}( - (\frac{{f_1^2}}{{b_1^4}} + \frac{{f_2^2}}{{b_2^4}} + \frac{{f_3^2}}{{b_3^4}}) + 2{\lambda_{10} })\\
\frac{{{{\ddot b}_1}}}{{{b_1}}} + \frac{{{{\dot b}^2}_1}}{{b_1^2}} + 3\frac{{\dot a}}{a}\frac{{{{\dot b}_1}}}{{{b_1}}} + 2\frac{{{{\dot b}_1}}}{{{b_1}}}(\frac{{{{\dot b}_2}}}{{{b_2}}} + \frac{{{{\dot b}_3}}}{{{b_3}}}) = \frac{1}{8}(7\frac{{f_1^2}}{{b_1^4}}-\frac{{f_2^2}}{{b_2^4}} - \frac{{f_3^2}}{{b_3^4}} + 2{\lambda_{10} }) - \frac{1}{{b_1^2}}\\
\frac{{{{\ddot b}_2}}}{{{b_2}}} + \frac{{{{\dot b}^2}_2}}{{b_2^2}} + 3\frac{{\dot a}}{a}\frac{{{{\dot b}_2}}}{{{b_2}}} + 2\frac{{{{\dot b}_2}}}{{{b_2}}}(\frac{{{{\dot b}_3}}}{{{b_3}}} + \frac{{{{\dot b}_1}}}{{{b_1}}}) = \frac{1}{8}(7\frac{{f_2^2}}{{b_2^4}}  -\frac{{f_3^2}}{{b_3^4}} -\frac{{f_1^2}}{{b_1^4}}+2{\lambda_{10} }) - \frac{1}{{b_2^2}}\\
\frac{{{{\ddot b}_3}}}{{{b_3}}} + \frac{{{{\dot b}^2}_3}}{{b_3^2}} + 3\frac{{\dot a}}{a}\frac{{{{\dot b}_3}}}{{{b_3}}} + 2\frac{{{{\dot b}_3}}}{{{b_3}}}(\frac{{{{\dot b}_1}}}{{{b_1}}} + \frac{{{{\dot b}_2}}}{{{b_2}}}) = \frac{1}{8}(7\frac{{f_3^2}}{{b_3^4}} -\frac{{f_1^2}}{{b_1^4}} -\frac{{f_2^2}}{{b_2^4}} + 2{\lambda_{10} }) - \frac{1}{{b_3^2}}\\
\end{array}\end{equation}
which describes a Kaluza-Klein cosmology with radii evolving in time and generalizes the ansatz of section 3 to separate radii and flux for each internal two sphere. Finally one can add the dilaton and dilaton potential to the system of equations from the Fischler-Susskind mechanism to obtain an early Universe string cosmology. When including the dilaton it is somewhat easier to use the dilaton action in the Einstein frame as in \cite{Poletti:1994ff}:
\begin{equation}S = \int {{d^{10}}x\sqrt { - g} } (\frac{1}{2}R - \frac{1}{4}{(\partial \phi )^2} - V(\phi ) - \frac{1}{4}{e^{ - \phi /2}}{F^2})\end{equation}
with dilaton potention $V(\phi)$. This leads to the equations:
$$\frac{1}{2}\square \phi  = \frac{{dV(\phi )}}{{d\phi }} - \frac{1}{4}{e^{ - \phi /2}}{F^2}$$
$$\nabla_\mu \left[ {\sqrt { - g} {e^{ - \phi /2}}{F^{\mu \nu }}} \right] = 0$$
\begin{equation}{R_{\mu \nu }} = \frac{1}{2}{\partial _\mu }\phi {\partial _\nu }\phi  + \frac{1}{4}{g_{\mu \nu }}V(\phi ) + {e^{ - \phi /2}}({F_{\mu \rho }}F_\nu ^\rho  - \frac{1}{{16}}{g_{\mu \nu }}{F^2})\end{equation}
which can be used together with the ansatz (7.9) to define a dilaton-radion-gauge cosmology. The dilaton potential $V(\phi) = g e^\phi$ was investigated in \cite{Akrami:2018ylq} and shown to be in disagreement with cosmological data. It would be interesting to investigate other nonperturbative dilaton potentials considered in the era of precision cosmology. Some of these potentials considered in the literature are listed in Table 5.

\begin{table}[h]
\begin{tabular}{|l|l|}
\hline
Potential       & Description   \\ \hline
$\lambda_{10} e^{\frac{5}{2}\phi}$   & Dilaton,   10d            \cite{Charmousis:2001nq}\cite{Charmousis:2009xr}\cite{Abdolrahimi:2011zg}\cite{Banerjee:2001ps}       \\ \hline
$ Ae^{4\phi}$   &       Dilaton, 4d               \cite{Horne:1993sy} \\ \hline
$e^{2\phi }{b^{ - 6}}({\lambda _{10}}{e^{2\phi }} - 3{b^{ - 2}})$            & Dilaton, radion, curvature                                          \cite{Obied:2018sgi}\cite{Akrami:2018ylq}\\ \hline
${e^{2\phi }}{b^{ - 6}}({\lambda _{10}}{e^{2\phi }} - 3{b^{ - 2}} + .5{f^2}{b^{ - 4}})$         &       Dilaton, radion, curvature, flux                                  \\ \hline
${e^{2\phi }}{b^{ - 6}}({\lambda _{10}}{e^{2\phi }} - 3{b^{ - 2}} + .5{f^2}{b^{ - 4}}) + .25{\mu ^2}{\sinh ^2}[2\phi ]$            & Modified dilaton, radion, curvature, flux                    \cite{Gregory:1992kr}                      \\ \hline
${e^{2\phi }}{b^{ - 6}}({\lambda _{10}}{e^{2\phi }} - 3{b^{ - 2}} + .5{f^2}{b^{ - 4}}) + {\mu ^2}{\phi ^2}$ & Modified dilaton, radion,
curvature, flux                  \cite{Gregory:1992kr}\cite{Horne:1992bi}                      \\ \hline
$\sum\nolimits_{i = 1}^s {{\lambda _i}{e^{4{g_i}\phi }}} $        & Dilaton, higher orders       \cite{Poletti:1994ff}\cite{Lim:2019hci}                                   \\ \hline
$ \exp ( - a{e^{ - 4\phi }})(A{e^{4\phi }} + B + C{e^{ - 4\phi }})$      & Dilaton, nonperturbative                   \cite{Poletti:1994ff}                       \\ \hline
\end{tabular}
\caption{\label{tab:table-name} Various dilaton-radion potentials that can be used to describe the cosmology for nonsupersymmetric string theory  
}
\end{table}

As the cosmological constant $\lambda_{10}$ is a loop effect it is interesting to include other loop effects such as the Casimir potential in the effective radion potential. This has been done for tori for the $SO(16)\times SO(16)'$ nonsupersymmetric theory in \cite{Ginsparg:1986wr}\cite{Nair:1986zn}\cite{Hamada:2015ria} and also for nonsupersymmetric orbifolds in \cite{Florakis:2016ani}. In the string frame one typically finds volcano type potentials vanishing at small and large $b$ with a stable dip ontop of the hill separating the too regions. For manifolds with spatial curvature such as $S^2 \times S^2 \times S^2$ one can form the sum over states to obtain a representation for the Casimir energy \cite{Birmingham:1988iv} but a full string computation has not yet been constructed.

\section{Quantum cosmology of the dimensionally reduced theory}

It is interesting to extend the cosmology discussion of section 4 to include quantum effects to form a quantum cosmology. For example one can consider a two Killing vector field reduction of these cosmologies on $S^2 \times S^2 \times S^2 \times S^2$ with an effective dimensional reduction to $1+1$ dimensions with ansatz:
\begin{equation}d{s^2} =  - d{t^2} + {a^2}(t,\theta)d\theta^2 + b^2_0(t,\theta)d\Omega _2^{(0)2} + b_1^2(t,\theta)d\Omega _2^{(1)2} + b_2^2(t,\theta)d\Omega _2^{(2)2} + b_3^2(t,\theta)d\Omega _2^{(3)2}\end{equation}
and as before we choose $b=b_i$ and fluxes $f=f_i$ for simplicity. The effective action reduces to \cite{Cadoni:1999gh}\cite{Cadoni:1994av}:
$$\int {{d^2}x} \sqrt { - g} {e^{ - 2\phi }}{b^8}\{ R + 4{(\nabla \phi )^2} + 56\frac{{{{(\nabla b)}^2}}}{{{b^2}}} - 32\nabla \phi  \cdot \frac{{\nabla b}}{b} + 8\frac{1}{{{b^2}}}$$
\begin{equation} - 2({\lambda _{10}}{e^{2\phi }} + V(\phi) +  \frac{{f^2}}{{{2b^{4}}}})\}\end{equation}
and defining $\varphi$ so the $e^{-2\varphi}=e^{-2\phi}b^8$ we have:
$$\int {{d^2}x} \sqrt { - g} {e^{ - 2\varphi }}\{ R + 4{(\nabla \varphi )^2} -8\frac{{{{(\nabla b)}^2}}}{{{b^2}}} + 8\frac{1}{{{b^2}}}$$
\begin{equation} - 2({\lambda _{10}}{e^{2\varphi }}{b^8} + V(\varphi +4 \log(b)) +  \frac{{f^2}}{{{2b^{4}}}})\}\end{equation}
This type of Lagrangian can be related to dilaton gravity in a minisuperspace with the equation of motion  in the form of quantized wave maps related to a sigma model target space  \cite{Tao}\cite{Chrstodoulou}\cite{Berger:1994sm}. In terms of the ansatz the zero mode portion of the action action becomes \cite{Berger:1972pg}\cite{Berger:1975kn}\cite{FernandoBarbero:2010qy}
\cite{McGuigan:1990nd}
:
\begin{equation}\int {dtN} {e^{ - 2\varphi }}a\{  - 4\frac{{{{\dot \varphi }^2}}}{{{N^2}}} + 4\frac{{\dot \varphi \dot a}}{{{N^2}a}} + 8\frac{{{{\dot b}^2}}}{{{b^2}{N^2}}} - 2( - \frac{4}{{{b^2}}} + {\lambda _{10}}{e^{2\varphi }}{b^8} + V(\varphi  + 4\log (b)) + \frac{{{f^2}}}{{2{b^4}}})\} \end{equation}
Variation with respect to $N$ yields the Hamiltonian constraint:
\begin{equation}{H_0} =  - 4\frac{{{{\dot \varphi }^2}}}{{{N^2}}} + 4\frac{{\dot \varphi \dot a}}{{{N^2}a}} + 8\frac{{{{\dot b}^2}}}{{{b^2}{N^2}}} + 2( - \frac{4}{{{b^2}}} + {\lambda _{10}}{e^{2\varphi }}{b^8} + V(\varphi  + 4\log (b)) + \frac{{{f^2}}}{{2{b^4}}})=0\end{equation}
and the canonical momentum are given by:
$$\frac{N}{a}a{\pi _a} = 4{e^{ - 2\varphi }}\dot \varphi $$
$$\frac{N}{a}{\pi _\varphi } =  - 8{e^{ - 2\varphi }}\dot \varphi  + 4{e^{ - 2\varphi }}\frac{{\dot a}}{a}$$
\begin{equation}\frac{N}{a}b{\pi _b} = 16{e^{ - 2\varphi }}\frac{{\dot b}}{b}\end{equation}
The Hamiltonian constraint is then:
\begin{equation}
{H_0} =  - \frac{{{{(a{\pi _a})}^2}{e^{4\varphi }}}}{{4{a^2}}} + \frac{{a{\pi _a}(2a{\pi _a} + {\pi _\varphi }){e^{4\varphi }}}}{{4{a^2}}} + \frac{{{{(b{\pi _b})}^2}{e^{4\varphi }}}}{{32{a^2}}} + 2( - \frac{4}{{{b^2}}} + {\lambda _{10}}{e^{2\varphi }}{b^8} + V(\varphi  + 4\log (b)) + \frac{{{f^2}}}{{2{b^4}}})=0
\end{equation}
The quantized Hamiltionan Wheeler DeWitt constraint is then of the form:
\begin{equation}H_{0}\Psi_{f}(a,b,\varphi) = 0\end{equation}
for states $ \Psi_{f}(a,b,\phi)$. The notion of superposition holds for this type of equation and for each solution we can form another solution:
\begin{equation}\Phi ({a},{b},{\varphi}) = \sum\limits_f {{A_{f} {\Psi _{f}({a},{b},{\varphi})}}}      \end{equation}
If there are transversable wormholes in the theory \cite{Gao:2016bin} as discussed in the context of two dimensional dilaton gravity in \cite{Maldacena:2017axo} one can also consider entangled states of the form
\begin{equation}\frac{1}{{\sqrt 2 }}({\Psi _{f}} \otimes \Psi {'_{f'}} \pm \Psi {'_{f'}} \otimes {\Psi _{f}})\end{equation}
where $ \Psi _{f}$ and $ \Psi'_{f'}$ are states on either side of the transversable wormhole.
Similar two dimensional dilaton gravity models have been recently discussed with the quantum gravity path integral with a sum over topologies as being related to matrix models \cite{Saad:2019lba}\cite{Cotler:2019nbi}\cite{Iliesiu:2019xuh}\cite{Arefeva:2019buu}. It would be of great interest to extend these methods to actions of the form of two dimensional reduced actions with additional scalar fields perhaps with more complicated matrix models. Also higher dimensional extensions of the matrix model methods have been discussed in terms of tensor models in \cite{Gurau:2011xp}\cite{Pereira:2019dbn}.

\section{Conclusion}

In this paper we discussed  the implications of the $SO(16) \times SO(16)'$ nonsupersymmetric model  for dark energy, dark matter, Higgs physics, and cosmology. The motivation for the study comes from the lack of evidence at the LHC for low energy supersymmetry so that nonsupersymmetric models need to be considered and one should examine more closely implications of theories of quantum gravity that have positive vacuum energy. In the $SO(16) \times SO(16)'$ nonsupersymmetric model dark matter can be represented by dark glueballs, a bifundamental fermion field can represent the connector or portal field connecting the visible sector to the dark matter. This can be important  for both accelerator searches for dark matter and astrophysical constraints as in both cases the portal field and its connection with the visible and hidden sectors is a driving factor in  experimental aspects of the model. For the Higgs field if it is represented as a an extra dimensional component of a gauge field this too could have important effects with regards to Higgs decay in the dark sector. For dark energy the  we showed that the compactified vacuum energy in four dimensions can be much reduced from the large one-loop value calculated in ten dimensions. Thus the extra dimensions and non-zero vacuum energy in ten dimensions play a joint role of driving spontaneous compactification and also providing a way to absorb the extra vacuum energy into the extra dimensions. This was not possible in supersymmetric theories in 10d and 11d as one considers large negative vacuum energy in four dimensions in those theories to exactly balance positive curvature and flux in the  extra dimensions. We also derived the Hamiltonian constraints for the theory with Kaluza Klein scalars as well as with gauge fields and discussed the behavior of the effective potential at large energies. Finally the consistency of the fundamental $SO(16) \times SO(16)'$ nonsupersymmetric model needs to be examined more closely especially with regards to realizing a positive value of the vacuum energy in the Fischler-Susskind mechanism and the swampland conjecture in string theory. In particular finding a way to remove the runaway potential of the dilaton through Casimir energy of the compactifield Horava-Witten theory or through fermion condensates would provide a way to stabilize the dilaton field, give it a mass, and provide a closer match to what we observe for low energy gravity.

\section*{Acknowledgement}
Michael McGuigan is supported from DOE HEP Office of Science de-sc0019139: Foundations of Quantum Computing for Gauge Theories and Quantum Gravity.

\end{document}